\documentclass[submission, Phys]{SciPost}
\usepackage{braket}
\usepackage{tikz}
\usetikzlibrary{positioning, calc, fit, arrows.meta}
\usepackage{amsmath,amsthm,amssymb}
\usepackage{bbold}
\usepackage{bm} 
\usepackage{graphicx} 
\usepackage{comment}

\usepackage[T1]{fontenc}
\hypersetup{
    colorlinks,
    linkcolor={blue},
    citecolor={blue},
    urlcolor={blue}
}

\usepackage{ulem}
\usepackage{bbm} 

\newcommand{\tr}[1]{\mbox{Tr}[{#1}]}

\begin{document}
\begin{center}{\Large \textbf{Replica Field Theory of Quantum Jumps Monitoring:\\ Application to the Ising Chain}
}\end{center}

\begin{center}
Youenn Le Gal\textsuperscript{1}, 
M. Schir\`o\textsuperscript{1}
\end{center}

\begin{center}
{\bf 1} JEIP, UAR 3573 CNRS, Coll\`{e}ge de France, PSL Research University, 11 Place Marcelin Berthelot, 75321 Paris Cedex 05, France\\
[\baselineskip]
\end{center}

\begin{center}
\today
\end{center}


\section*{Abstract}
{\bf
In this work we derive the replica field theory for monitored quantum many-body systems evolving under the quantum jumps protocol, corresponding to a non-Hermitian evolution interspersed with random quantum jumps whose distribution is state-dependent.
We show that the density matrix of $R$ replicas evolves according to a master equation where the non-Hermitian term is replica-diagonal while coupling among replicas are due to quantum jumps. We write down the associated Keldysh action and study its behavior for the specific case of the Ising Chain with monitoring of particle density and tunable anisotropy, interpolating between free fermions with strong U(1) symmetry and the Ising chain with Z$_2$ symmetry. We derive the effective field theory in terms of slowly varying fields and obtain the replica-diagonal saddle point, which we show to describe the average state. We then go beyond saddle point and derive the effective field theory describing the replica off-diagonal sector, which takes the form of a Non-Linear Sigma Model. The symmetry class is either DIII or D, depending on the parameters of the Ising chain, except at a special symmetric point, where we recover the results for free fermions. We discuss the implications of these findings for the entangling phase observed numerically for the monitored Ising chain.}

\vspace{10pt}
\noindent\rule{\textwidth}{1pt}
\tableofcontents
\thispagestyle{fancy}
\noindent\rule{\textwidth}{1pt}
\vspace{10pt}

\section{Introduction}
\label{sec:introduction}


The effect of quantum measurements on many-body systems have recently gained large attention, in particular in the context of monitored dynamics where unitary evolution is interrupted by strong or weak measurements. The random sequence of measurement outcomes, occurring with a probability set by the Born rule, defines a quantum trajectory evolving according a stochastic quantum dynamics. A major theme of recent research has been to characterize the properties of the system along the conditional evolution
given by a quantum trajectory, in particular its entanglement content. Indeed, while unitary dynamics generally leads to a growth of the entanglement entropy in time, the effect of non-unitary processes such as measurements or monitoring can slow down, or even halt the spread of quantum information, leading to a so called Measurement-Induced Phase Transition (MIPT)~\cite{liQuantumZenoEffect2018,liMeasurementdrivenEntanglementTransition2019,skinnerMeasurementInducedPhaseTransitions2019}. This manifests clearly in an entanglement transition
between a volume law phase for low measurement rates, when unitary dynamics dominates, to an area law phase above a threshold measurement rate, where entanglement becomes short-ranged due to the Quantum Zeno Effect.

The basic phenomenology of MIPT has been extensively studied in the context of random quantum circuits with projective measurements, see Ref.~\cite{fisherRandomQuantumCircuits2023} for a review, where efficient numerical simulations are available based on the Clifford group structure and analytical progress has been obtained by mapping the problem to effective statistical mechanics models. Monitored systems with deterministic unitary dynamics driven by an interacting many-body Hamiltonian are also expected to display a similar MIPT phenomenology, yet its understanding is less developed and to a large extent based on numerical simulations~\cite{fujiMeasurementinducedQuantumCriticality2020,xingInteractionsIntegrabilityWeakly2024}. The monitored dynamics in this context can be generated by projective measurements or by a non-unitary unravelling of the Lindblad master equation~\cite{wisemanQuantumMeasurementControl2009,daleyQuantumTrajectoriesOpen2014,fazioManyBodyOpenQuantum2024}, such as the Quantum-State Diffusion (QSD)~\cite{gisinQuantumstateDiffusionModel1992} or the Quantum Jumps (QJs) protocol~\cite{dalibardWavefunctionApproachDissipative1992,plenioQuantumjumpApproachDissipative1998}. 

Monitored Gaussian systems, in particular fermionic models with particle density monitoring, have been extensively studied numerically taking advantage of the Gaussianity of the state at the level of quantum trajectories. In particular, the monitored tight-binding chain with strong U(1) symmetry has since become a central example for investigating MIPTs using a variety of monitoring protocols, sparking a debate over whether a genuine phase transition occurs in this model in the thermodynamic limit~\cite{caoEntanglementFermionChain2019,albertonEntanglementTransitionMonitored2021,coppolaGrowthEntanglementEntropy2022}.
The monitored Ising chain has also been the focus of a large attention~\cite{biellaManyBodyQuantumZeno2021, turkeshiMeasurementinducedEntanglementTransitions2021,turkeshiEntanglementTransitionsStochastic2022,paviglianitiMultipartiteEntanglementMeasurementinduced2023,piccittoEntanglementTransitionsQuantum2022}. Here numerical evidence points towards the existence of a MIPT between a sub-volume phase and an area-law phase~\cite{legalEntanglementDynamicsMonitored2024}, a result which is known to persist also in the no-click limit of purely non-Hermitian evolution~\cite{turkeshiEntanglementCorrelationSpreading2023}. 

Despite recent progress in the numerical solution of stochastic many-body Schrodinger equation~\cite{fan2025entanglementdynamicsmonitorednoninteracting}, analytical and field theory approaches to MIPT are particularly valuable to gain further insights on the large scale behavior of monitored systems. However, the MIPT appears in non-linear functions of the state, such as the entanglement entropy or the purity~\cite{gullansDynamicalPurificationPhase2020}. Handling the average over the quantum measurement outcomes poses several challenges from the theoretical point of view. In order to handle the average over the measurement noise one can use the replica trick, as done in other classes of disordered problems in statistical physics. This amounts to replicating the system by introducing $R$ identical sets of degrees of freedom, all experiencing the same measurement noise.  Usually, in the replica method for disordered systems to obtain ultimately a physical quantity we must take the limit $R \rightarrow 0$, corresponding to averaging the free-energy of the system, i.e. the logarithm of the partition function. For monitored systems it was shown, both in random circuits~\cite{jianMeasurementinducedCriticalityRandom2020,baoTheoryPhaseTransition2020} and in monitored fermionic systems~\cite{poboikoNonlinearSigmaModels2023,jianMeasurementinducedEntanglementTransitions2023,favaNonlinearSigmaModels2023}, that the relevant replica limit corresponds to sending $R \rightarrow 1$ to account for the correct Born's rule in the weight of each quantum trajectory. In this framework, the unconditional (average state) dynamics corresponds to the replica-symmetric sector, whereas the conditional dynamics is characterized by the so-called replicon manifold, i.e. the subspace transverse to the symmetric one.

In the context of monitored free fermions, first attempts to a field theory descriptions either focused on the averaged state~\cite{yangKeldyshNonlinearSigma2023} or considered the dynamics in a sector at fixed number of replicas~\cite{buchholdEffectiveTheoryMeasurementInduced2021,baoSymmetryEnrichedPhases2021}. The replica field theory in the appropriate replica limit $R\rightarrow 1$ was discussed for free fermions with projective density measurements and strong U(1) symmetry in Ref.~\cite{poboikoNonlinearSigmaModels2023,poboikoMeasurementInducedPhaseTransition2024} using the Keldysh framework, which is natural to describe out of equilibrium phenomena and Anderson localization~\cite{kamenevFieldTheoryNonEquilibrium2011}. This leads to a Non-Linear Sigma Model (NLSM) description corresponding to the BDI class in the Altland-Zirbauer ten-fold way classification~\cite{altlandNonstandardSymmetryClasses1997}.
A similar approach to free fermions was extended to QSD monitoring in Ref.~\cite{chahineEntanglementPhasesLocalization2024} and later extended to interacting fermions in Ref.~\cite{poboikoMeasurementinducedTransitionsInteracting2025,guoFieldTheoryMonitored2024} and to fermionic Gaussian circuits with U(1) symmetry in Ref.~\cite{favaMonitoredFermionsConserved2024}. 
The effect of breaking the strong $U(1)$ symmetry to a weak one, i.e. in presence of particle gain and losses, has been discussed in Ref.~\cite{starchlGeneralizedZenoEffect2024}, and shown to not change qualitatively the physics. The role of the unravelling protocol, in particular interpolating between monitoring and classical noise, has been discussed recently for free-fermions in Ref.~\cite{niederegger2025absencemeasurementunravelinginducedentanglement}.
Field theory for monitored noisy Majorana fermions was discussed in Ref.~\cite{jianMeasurementinducedEntanglementTransitions2023,favaNonlinearSigmaModels2023,Tiutiakina2025fieldtheory}. In the case of Gaussian random circuits and under the quantum state diffusion protocol a NLSM was derived in Ref.~\cite{favaNonlinearSigmaModels2023} by mapping the monitored random circuit to a spin chain. The outcome of those studies is that the critical behavior of monitored free fermions depends strongly from the underlying symmetry: in the U(1) case a phenomenon similar to weak-localization correction makes the weak monitoring phase unstable and ultimately leads to an area-law phase, although above a crossover scale which is exponentially large in the coupling. For noisy Majorana fermions it was shown that an opposite mechanism is at play, anti-localization, protecting the weak-monitoring phase.

In this work we derive the Keldysh field theory for monitored systems under the quantum jumps protocol. Quantum Jumps are peculiar since their randomness is state-dependent, which poses some challenges for the analytical description. Another interesting aspect is the role of the non-Hermitian evolution in between quantum jumps for the MIPT, a topic which has attracted attention and which a field theory approach can help elucidating. Starting from the QJs protocol, we first derive an effective master equation for the replicated density matrix, and then write down the Keldysh action. We obtain a replica field theory describing $R$ replicas evolving with a non-Hermitian Hamiltonian which is replica-diagonal, the only coupling between replicas arises because of quantum jumps. To study in detail this field theory we focus on  MIPT in the monitored transverse field Ising chain with tunable anisotropy, interpolating between free fermions and the full Ising model. Despite significant numerical studies, this model remains, to our knowledge, largely unexplored in the field-theoretical treatment of monitored systems. 

We derive the effective field theory in terms of slowly varying space-time fields described by $4R\times 4R$ matrices in Keldysh, Nambu (or particle-hole) and replica space. We obtain the replica symmetric saddle point solution, and show that it coincides as expected with the results obtained via Lindblad dynamics. We pay particular attention to identifying properly the symmetries of the Keldysh action, and of the saddle point: we show that for the Ising chain with density monitoring the relevant symmetry is $SO(R)$ and the NLSM corresponds to the DIII symmetry class, except at two special points in the phase diagram, the so called pairing limit of the Ising chain which corresponds to the D class, and the fully isotropic limit which reduces to free-fermions and the AIII class discussed before. We then derive the appropriate NLSM describing the long-wavelength fluctuations associated to gapless Goldstone modes.  Finally, using the results obtained, we comment on the stability of the weak-monitoring phase and of the MIPT.

This manuscript is organized as follows. In Section \ref{sec:replica_frame}, we discuss how to handle conditional averages of non-linear observables via the replica trick. In Section \ref{sec:keldysh_frame} we use this approach to write down the replicated Keldysh field theory associated to monitored systems under QJs protocol. In Section \ref{sec:ising_app_keldysh}, we specialize our discussion to the monitored Ising chain for which we write down the replicated Keldysh action. In Section \ref{sec:NLSM_Ising} we derive the effective field theory which we study via saddle point in Section \ref{sec:ising_saddle}. In Section \ref{sec:symmetries} we present an analysis of the symmetries of the replicated theory, which then allows us to derive in Section \ref{sec:non_liner_sigma} the long-wavelength theory taking the form of a NLSM. Finally, in Section \ref{sec:discussion} we discuss our results, and present our conclusions in Section \ref{sec:conclusions}. Several technical Appendices complete this work.


\begin{figure*}[t]
    \centering
    \includegraphics[width=\textwidth]{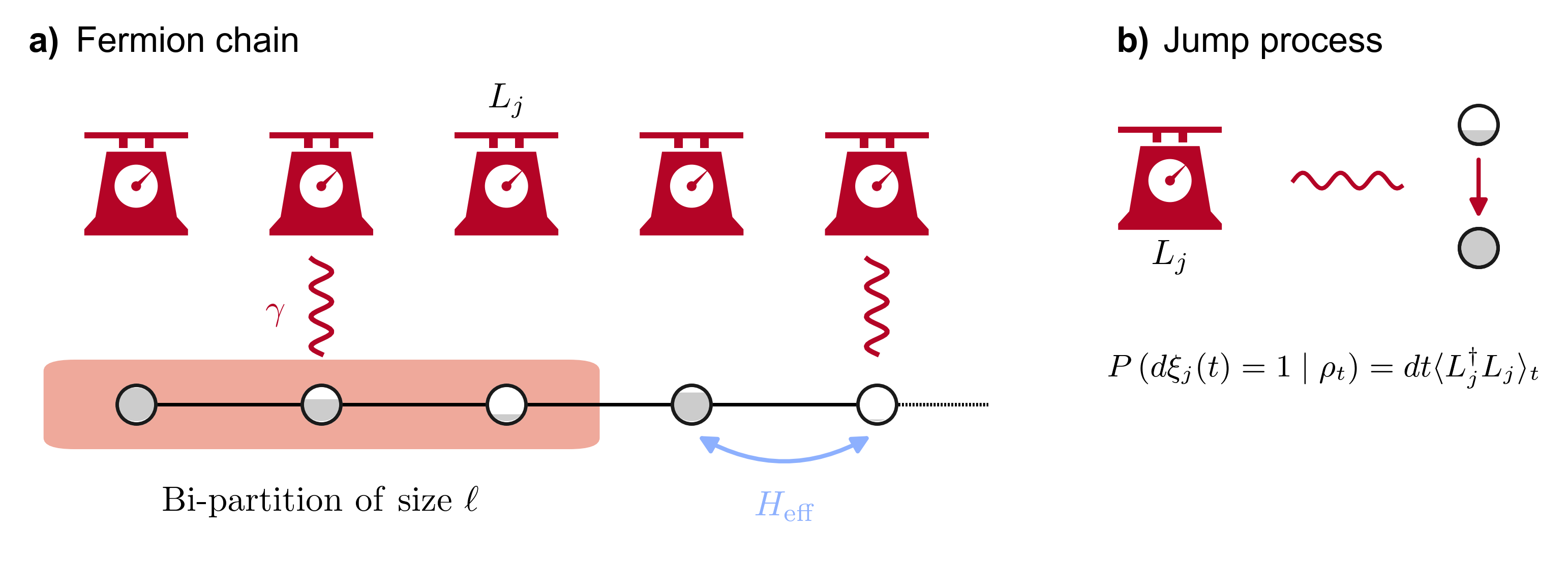}
\caption{Cartoon of the setup. A monitored fermionic chain evolving under quantum jump protocol, characterised by a deterministic evolution driven by a non-Hermitian Hamiltonian $H_{\rm nH}$ $a)$ and stochastic Quantum Jumps $b)$. }
\label{fig:non_interacting}
\end{figure*}

\section{General Framework for Monitoring by Quantum Jumps}
\label{sec:replica_frame}

In this Section we introduce the Quantum Jump monitoring protocol, discuss the structure of a quantum trajectory, the unconditional dynamics leading to a Lindblad master equation as well as how to deal with conditional averages using the replica trick. For concreteness, we consider a one dimensional lattice model with with Quantum Jumps (QJs) monitoring protocol, see Fig.~\ref{fig:non_interacting}, even though the idea of the replica trick is general, and holds also for higher dimensional systems and different monitoring protocols.


\subsection{Structure of a Quantum Trajectory}

In this work, we consider the QJs monitoring protocol, which is usually described by a Stochastic Schrödinger Equation (SSE) \cite{dalibardWavefunctionApproachDissipative1992,plenioQuantumjumpApproachDissipative1998,wisemanQuantumMeasurementControl2009,daleyQuantumTrajectoriesOpen2014}. In this protocol, each time step involves either the detection of a bath excitation—resulting in a quantum jump—or the absence of any excitation, which leads to a non-Hermitian evolution. Specifically, we focus on the situation where all channels are monitored with an identical measurement rate 
$\gamma$, resulting in the following conditional evolution for the state of the system $\vert\psi(t)\rangle$
\begin{equation}
    \ket{\psi(t+dt)} = \left(1 - \sum_{j=1}^L d \xi_j (t) \right) \frac{K_0}{\sqrt{\left\langle K_0^\dagger K_0 \right\rangle }} \ket{\psi(t)}  + \sum_{j=1}^L d \xi_j (t) \frac{K_0}{\sqrt{\left\langle K_j^\dagger K_j \right\rangle }}\ket{\psi(t)} ,
    \label{eq:sse_state}
\end{equation}
where $K_j$ are Kraus operators of a POVM fulfilling $\sum_{j=0}^L K_j^\dagger K_j = \mathbbm{1}$ to preserve probability normalisation, and $d\xi_j(t) = 0,1$ is a random variable called a Poisson increment, with $P\left( d\xi_j(t) = 1 \mid \rho_t \right) =   \tr{K_j^\dagger K_j \rho_t} $ where we observe that the process is conditional to the state $\rho_t$. 
These random variables take mostly the value zero, leading to non Hermitian evolution, and only occasionally take the value $d\xi_{x_m}(t_m) = 1$ which corresponds to a quantum jump occurring at time $t_m$ and in channel $x_m$\footnote{We note that we should write $\prod_{j=1}^L (1 - d\xi_j(t))$ in front of the first term of Eq.~\eqref{eq:sse_state} since it corresponds to the probability of no-excitation in any channel, but since $d\xi_i(t) \propto  dt$ in the limit where $dt$ tends toward $0$, it is equivalent to directly write $1 - \sum_{j=1}^L d\xi_j(t)$ }. The operator $K_0$ is taken to be close to the identity, reading as $K_0 = \mathbbm{1} - (J/2 + i H)dt$  where $H$ is the Hamiltonian of the system, while $Jdt =  \sum_{j=1}^L K_{j}^\dagger K_j $. The channels with $j \geq 1$ describe jumps such that $K_j = \sqrt{\gamma dt} L_j $. As a result, the evolution in between quantum jumps is dictated by the non-Hermitian Hamiltonian
\begin{align}\label{eqn:HnH}
H_{\rm nH}=H-iJ/2=H-i\frac{\gamma}{2}\sum_{j=1}^L L^{\dagger}_jL_j.
\end{align}

Hence, a quantum trajectory is a succession of non-Hermitian evolution and punctual jumps events, and we can write the resulting pure state (we consider only initial pure state) which underwent $M$ QJs between time $t_i$ and $t_f$ as 
\begin{equation}\label{eq:psitilde}
    \ket{\tilde{\psi}(t_f,\mathcal{T})} = U_\mathrm{nH}(t_f,t_M) L_{x_M}  U_\mathrm{nH}(t_M,t_{M-1}) K_{x_{M-1}} ...   K_{x_{1}} U_\mathrm{nH}(t_1,t_i) \ket{\psi_0},
\end{equation}
where $\ket{\psi_0}$ is the initial pure state, $ U_\mathrm{nH}(t_b,t_a)=e^{-iH_{\rm nH}(t_b-t_a)}$ is the non-Hermitian evolution operator between time $t_a$ and $t_b$, $t_m$ (resp. $x_m$) is the time (resp. the jump channel) of the $m^{th}$ jumps. The Kraus operators $K_j$ for $1 \leq j \leq L$ correspond to jumps channels. We call $\mathcal{T} = \left\{(x_1, t_1), ... , (x_M , t_M) \right\}$ the set of different times and channels of the jumps happening during the trajectory up to time $t$, which completely define a quantum trajectory. The symbol $\tilde{\circ }$ indicates that the state in Eq.~\eqref{eq:psitilde} is not normalized,  since non-unitary operations are not preserving the norm. To obtain a valid quantum state we must normalize it, resulting in  
\begin{equation}
    \ket{\psi(t_f,\mathcal{T})} =    \ket{\tilde{\psi}(t_f,\mathcal{T})}/\sqrt{\mbox{Tr}[\tilde{\rho}_{t,\mathcal{T}} }]\quad \mathrm{and} \quad  \rho_{t_f,\mathcal{T}} =    \tilde{\rho}_{t_f,\mathcal{T}}/ \mbox{Tr}[\tilde{\rho}_{t_f,\mathcal{T}} ] ,
\end{equation}  
where $\tilde{\rho}_{t_f,\mathcal{T}} = \ket{\tilde{\psi}(t_f,\mathcal{T})} \bra{\tilde{\psi}(t_f,\mathcal{T})}$.  The probability $P\bigl[\mathcal{T} \bigr]$ to obtain a given trajectory, which depends on the different waiting times (time between jumps) and jump channels, is thus simply given by the Born rule, which coincides with the norm of the unnormalised state, reading as 
\begin{equation}
    P\bigl[ \mathcal{T} \bigr] = \tr{\tilde{\rho}_{t_f,\mathcal{T}}} .
\end{equation}

We note that the probability $P_\mathrm{WT}\left(\tau, \rho_{t_m}\right)$ to wait a time $\tau= t_{m+1} - t_{m}$ between two jumps is given by the decay of the norm due to the non Hermitian evolution, in particular $P_\mathrm{WT} \left(\tau, \rho_{t_m}\right) = \left \langle U_\mathrm{nH}^\dagger(t_{m+1}, t_{m}) U_\mathrm{nH}(t_{m+1}, t_{m}) \right\rangle_{t_{m}}$, and the conditional probability $P_\mathrm{qj}\left(x_m, \rho_{t_m} \right)$, to jump in channel $x_m$ knowing that a jump happens at $t_m$, is given by $P_\mathrm{qj}\left(x_m, \rho_{t_m} \right) = \left\langle L_{x_m}^\dagger L_{x_m} \right\rangle_{t_m}$.


\subsection{Conditional Averages and the Replica Trick}

To capture the physics of MIPT, encoded as discussed in the Introduction in non-linear functionals of the conditional density matrix, we will be interested in computing higher-order moments of $\rho_{t_f,\mathcal{T}}$, and then perform the average over the measurement outcomes. This reads for the case of the $N^\mathrm{th}$ moment of an observable $Q$ as 
\begin{equation}
    \overline{\left\langle Q \right\rangle_\mathcal{T}^N} = \int_{\mathcal{T}}   P\bigl[ \mathcal{T} \bigr]  \left\langle Q \right\rangle_\mathcal{T}^N ,
\end{equation}
where $\int_{\mathcal{T}}$ stands for the summation over all the trajectories which can be expressed as  $\int_{\mathcal{T}}=\sum_{m=0}^\infty \prod_{m=1}^M \left(  \sum_{x_m=1}^L \int_{t_{m-1}}^{t_f} dt_m \right)$.  Interestingly, we can rewrite this quantity in a linear manner by replicating $N$ times the density matrix 
\begin{align}
    \overline{\left\langle Q \right\rangle_\mathcal{T}^N} =  \int_{\mathcal{T}}   P\bigl[ \mathcal{T} \bigr] \mathrm{Tr}\left[ Q^{\otimes N} \rho_{t_f}^{\otimes N} \right]   =   \mathrm{Tr}\left[ Q^{\otimes N} \overline{\rho_{t_f}^{\otimes N}} \right],   
\end{align}
with the trace that now acts in a $N$ replicated space. Therefore the object of interest is $\overline{\rho_{t_f}^{\otimes N}}$. \\

Still, because of the measurements, the density matrix needs to be renormalised as $\rho_{t_f,\mathcal{T}} = \tilde{\rho}_{t_f,\mathcal{T}} / \mathrm{Tr}[\tilde{\rho}_{t_f,\mathcal{T}}]$, leading to a typical average of disordered systems
\begin{equation}
    \overline{\rho_{t_f}^{\otimes N}} = \int_{\mathcal{T}}   P\bigl[ \mathcal{T} \bigr] \left( \rho_{t_f,\mathcal{T}} \right)^{\otimes N} =  \int_{\mathcal{T}}   P\bigl[ \mathcal{T} \bigr] \frac{\left( \tilde{\rho}_{t_f,\mathcal{T}} \right)^{\otimes N}}{\mathrm{Tr}[\tilde{\rho}_{t_f,\mathcal{T}}]^N } .
    \label{eq:average_traj}
\end{equation}
We thus use a replica trick, to rewrite the denominator as a trace by introducing $R \geq N$ replicas of the unormalised density matrix, and perform a partial trace over $R-N$ of them\footnote{Indeed we have $\mathrm{Tr}_{r=N+1, ...,R} \left[ \prod_{r=1}^{R} \tilde{\rho}_{t_f,\mathcal{T}}^{(r)} \right] = \tr{\tilde{\rho}_{t_f,\mathcal{T}}}^{R-N} \tilde{\rho}_{t_f,\mathcal{T}}^{N}$ }, i.e.
\begin{equation}
    \overline{\rho_{t_f}^{\otimes N}} =  \int_{\mathcal{T}}  \tr{\tilde{\rho}_{t_f,\mathcal{T}}}^{1-N}  \left( \tilde{\rho}_{t_f,\mathcal{T}} \right)^{\otimes N} = \lim_{R \rightarrow 1} \int_{\mathcal{T}} \mathrm{Tr}_{r=N+1, ...,R} \left[ \prod_{r=1}^{R} \tilde{\rho}_{t_f,\mathcal{T}}^{(r)} \right],
    \label{eq:average_traj_replica}
\end{equation}
where $\tilde{\rho}_{t_f,\mathcal{T}}^{(r)}$ is the $r^{th}$ unormalised replicated density matrix, where for any operator $X$ we define its replicated version $X^{(r)} =\left(\bigotimes_{\alpha=1}^{r-1} \mathbbm{1} \right) \otimes X \otimes \left(  \bigotimes_{\alpha=1}^{R-r} \mathbbm{1}  \right)$. 
When taking the limit $R\rightarrow1$, we obtain the correct expression for the average, and it is because each trajectory has a different weight $P\bigl[ \mathcal{T} \bigr]$ due to Born's rule, that the unusual limit $R\rightarrow 1$ appears. In the following $\overline{\square}$ will always denote $\overline{\square} = \int_{\mathcal{T}} \square = \mathbbm{E}\left[ \square \right]$, where we can notice the absence of $ P\bigl[ \mathcal{T} \bigr]$, since it is included through the partial trace and the limit $R\rightarrow 1$. This also means that $\overline{\rho}$ is un-normalised until the limit $R\rightarrow 1$ is taken, and for conciseness and readability we write $\rho_{t_f}^N = \overline{\rho_{t_f}^{\otimes N}}$.

To summarize, the replica trick applied to monitored quantum systems allows to handle the averaging over the measurement noise, by simply considering $R$ copies of unnormalised density matrix of the system averaged over the noise and taking the limit $R\rightarrow 1$ limit in the end, to restore the correct Born rule weight. 

\section{Keldysh Field theory for Quantum Jumps Protocol}
\label{sec:keldysh_frame}

In this Section, we write down the field theory associated to the dynamics of the replicated density matrix, thus effectively converting the usual QJs protocol into a field theory that can be treated with tools from many-body physics. Since the dynamics of monitored systems is in general out of equilibrium, we employ the Keldysh formalism developed for such cases. General reviews of this formalism are available \cite{kamenevFieldTheoryNonEquilibrium2011,siebererKeldyshFieldTheory2016,siebererUniversalityDrivenOpen2025}, and some important features of our case are detailed in Appendices \ref{app:regularisation_green_function} and \ref{app:causality_struct}. 
We start by writing down a stochastic equation for the $N-$replicated density matrix of the system, and then average over the noise to obtain a master equation, similarly to what is usually done in the context of Lindblad equation \cite{landiCurrentFluctuationsOpen2024}. From this master equation it is then easy to find the corresponding Keldysh action.
 
\subsection{Replicated Stochastic Equation and Master Equation}
We start from the Stochastic Schrödinger Equation (SSE) Eq.~\eqref{eq:sse_state} previously introduced and even if the purity of the state is conserved, for convenience we rewrite this equation in terms of the density matrix 
\begin{equation}
    \rho_{t+dt} = \frac{ 1 - \sum_{j=1}^{L} d\xi_j (t) }{\prod_{j=1}^L P \Bigl( d\xi_j(t) = 0 \mid \rho_t \Bigr)} K_0 \rho_t K_0^\dagger + \sum_{j=1}^L \frac{ d \xi_j(t) }{P\Bigl(d\xi_j(t) = 1 \mid \rho_t \Bigr)} K_j \rho_t K_j^\dagger.
    \label{eq:master_eq_povm}
\end{equation}
We now want to write the $N$-replicated version of Eq.~\eqref{eq:master_eq_povm}. We can proceed by replicating each term independently: indeed cross products between channel $0$ and channels $j \geq 1$ are avoided, since if one $d\xi_j(t) = 1$ then $1 - \sum_{j=1}^{L} d\xi_j (t) = 0$, and among the channels $j \geq 1$ since these terms are of order $dt^2$. We thus obtain 
\begin{equation}
    \rho_{t+dt}^{\otimes N} = \frac{1 - \sum_{j=1}^{L} d\xi_j (t)}{p_\mathrm{nH}^N}   \left[ K_0 \rho_t K_0^\dagger \right]^{\otimes N} + \sum_{j=1}^L \frac{d \xi_j(t) }{ \left(p_\mathrm{qj}^{(j)} \right)^N   } \left[ K_j \rho_t K_j^\dagger \right]^{\otimes N },
\end{equation}
where $p_\mathrm{nH} = \prod_{j=1}^L P\Bigl( d\xi_j(t) = 0 \mid \rho_t \Bigr)$ and $p_\mathrm{qj}^{(j)} =  P\Bigl(d\xi_j(t) = 1 \mid \rho_t \Bigr)$. By noticing that $ p_\mathrm{nH}  \underset{dt \to 0}{=} 1 - \sum_{j=1}^L  P\Bigl(d\xi_j(t) = 0 \mid \rho_t \Bigr) =  \tr{K_0 \rho_t K_0^\dagger  } $, we find using the replica trick that
\begin{equation}
    \rho_{t+dt}^{\otimes N}  = \lim_{R\rightarrow 1 } \mathrm{Tr}_{N+1,..., R }\left[  \frac{1 - \sum_{j=1}^{L} d\xi_j (t)}{p_\mathrm{nH}}    \left[ K_0 \rho_t K_0^\dagger \right]^{\otimes R}  + \sum_{j=1}^L \frac{d \xi_j(t)}{p_\mathrm{qj}^{(j)} }  \left[ K_j \rho_t K_j^\dagger \right]^{\otimes R }  \right].
    \label{eq:replicated_channel}
\end{equation}
In the following, we hence consider the $R$-replicated density matrix and the limit $R \rightarrow 1$ should only be taken at the very end. 
We now average Eq.~\eqref{eq:replicated_channel}, and use that $\mathbbm{E} \Bigl[ d\xi_j (t) f[\rho ](t)  \mid \rho_t \Bigr] = \mathbbm{E} \Bigl[ d\xi_j (t)  \mid \rho_t \Bigr] f[\rho ](t)  =  P\Bigl(d\xi_j(t) = 1 \mid \rho_t\Bigr) f[\rho ](t) $, where $f$ is any function of the density matrix \cite{landiCurrentFluctuationsOpen2024}. By doing so, prefactors in front of the different terms are canceling, and we obtain the following replicated master equation \footnote{To be precise, here we are proceeding like in the usual derivation of the Lindblad equation, from the stochastic quantum jumps equation. Specifically we have 
\begin{equation}
   \nonumber \mathbbm{E} \left[ \left( d\xi_j(t)/  p_\mathrm{qj}^{(j)}    \right) f\left[ \rho \right] \right] = \mathbbm{E}\left[  \mathbbm{E}\left[   \left( d\xi_j(t)/  p_\mathrm{qj}^{(j)}   \right)  f\left[ \rho \right] \mid \rho_t \right] \right] = \mathbbm{E}\left[     \left( p_\mathrm{qj}^{(j)} /  p_\mathrm{qj}^{(j)}   \right)  f\left[ \rho\right]   \right]   = f(\overline{\rho}).
\end{equation}
We stress that the cancellation is necessary since otherwise $p_{\mathrm{qj}}^{(i)}$ generally depends on the state, and we cannot go further in the averaging. 
}
\begin{equation}
    \rho^R_{t+dt}  =    \left( \mathbbm{1}_R - i dt H_\mathrm{nH}^R\right)  \rho^R_{t} \left( \mathbbm{1}_R - i dt H_\mathrm{nH}^R\right) ^\dagger + \gamma dt  \sum_{j=1}^L   L_j^R \rho^R_{t} \left( L_j^R \right) ^\dagger ,  
    \label{eq:replicated_master_eq}
\end{equation}
where $H_\mathrm{nH}$ is defined in Eq.~(\ref{eqn:HnH}) and we have introduced the shortcut notation $\rho^R = \overline{\rho^{\otimes R}}$ and $X^R = \prod_{r=1}^R X^{(r)}$ . 
Few comments are in order here: first we see that for $R=1$ this equation reduces to a Linbdlad master equation, as expected since in this case we are just averaging the conditional density matrix. On the other hand, for $R>1$ the un-normalised replicated density matrix evolves with an equation that has not a Lindblad form. 
In particular, the non-Hermitian evolution (arising from $H_\mathrm{nH}^R$) is independent for each replica $r$, which are then coupled by the quantum jump term, the last term in Eq.~(\ref{eq:replicated_master_eq}). The latter involves a highly non-linear term with the product of $R$ jump operators. As a result of the structure of the back-action non-Hermitian evolution and quantum jumps term, the map in Eq.~(\ref{eq:replicated_master_eq}) does preserve the trace of the density matrix only in the replica limit $R\rightarrow1$. From this master equation, we can obtain the field theory following the standard procedure presented for example in \cite{siebererKeldyshFieldTheory2016}. It is also possible to derive the master equation in Eq.~(\ref{eq:replicated_master_eq}) in a time integrated way as presented in Appendix \ref{app:time_int_master_eq}.

\subsection{The General Form of the Keldysh Action}

Having obtained a master equation for the replicated density matrix in Eq.~\eqref{eq:replicated_master_eq}, it is now a standard exercise to write the associated Keldysh action over replicated fermionic coherent states, $\psi_{\pm,r}$  and $\overline{\psi}_{\pm,r}$, where $r$ labels the replica and $\pm$ the branchs of the Keldysh contour \cite{siebererKeldyshFieldTheory2016,siebererUniversalityDrivenOpen2025,kamenevFieldTheoryNonEquilibrium2011}. In particular, we can write 
\begin{align}
\mathrm{Tr}\left[ \rho^R \right] =\int D \psi D\bar{\psi}e^{iS},
\end{align} 
with the Keldysh action 
\begin{align}
    S= S_\mathrm{nH} + S_\mathrm{QJ},
\end{align}  
made up of two contributions. The first term  $S_\mathrm{nH}$ describes the Keldysh action associated to the non-Hermitian evolution of the $R$ replicated fields and it is hence replica diagonal
\begin{multline}
    S_\mathrm{eff} = \int_{t_i}^{t_f} dt \sum_{r=1}^R \left[\sum_{x=1}^L 
    (i \overline{\psi}_{x,+,r} \partial_t \psi_{x,+,r} - i \overline{\psi}_{x,-,r} \partial_t \psi_{x,-,r} )  - \left[  H_{r,+} - H_{r,-} \right] \right. \\ \left. +i \frac{\gamma}{2} \sum_{x=1}^L  L^\dagger_{x,r,+}L_{x,r,+} + L^\dagger_{x,r,-} L_{x,r,-}   \right], 
    \label{eq:action_gen_eff}
\end{multline}
where $H_{r,\pm} = H(\overline{\psi}_{\pm,r},\psi_{\pm,r})$ and  $L_{x,r,\pm} = L_x(\overline{\psi}_{\pm,r},\psi_{\pm,r})$. The second term $S_\mathrm{QJ}$ accounts for the effect of quantum jumps and reads therefore as
\begin{equation}
    S_\mathrm{QJ} = -i \gamma \int_{t_i}^{t_f}dt\sum_{x=1}^L \prod_{r=1}^R L_{x,r,+} L^\dagger_{x,r,-}.
    \label{eq:action_gen_meas}
\end{equation}
We recall that as usual in continuous Keldysh formalism, the initial state dependence is not explicitly written in the action since it is a boundary term, however it should be taken into account in the Keldysh component of the Green's functions.
For $R=1$, i.e. a single replica corresponding to averaging the conditional density matrix, the result above reduces to the usual
Lindblad action \cite{siebererKeldyshFieldTheory2016,thompsonFieldTheoryManybody2023}. However for generic $R>1$ the replicated action differs from a conventional Keldysh action of nonequilibrium problems since it does not preserve the normalisation of the density matrix, a consequence of Eq.~\eqref{eq:replicated_master_eq} not being in Lindblad form. We are going to come back to this point in the next section, when discussing the case of the Ising chain.

\section{The Case of the Monitored Ising chain }
\label{sec:ising_app_keldysh}
 
We now specify the discussion with an explicit model, the monitored one dimensional Ising chain, whose Hamiltonian reads as 
\begin{equation}\label{eq:IsingChain}
    H = \sum_{j=1}^L J_x \sigma_{j+1}^x \sigma_j^x + J_y \sigma_{j+1}^y \sigma_j^y + h \sigma_j^z,
\end{equation}
where $L$ is the size of the chain and $\sigma^\alpha_j$ are Pauli matrices ($\alpha \in \{x, y,z \}$) acting on site $j$ of the chain. Here $J_x,J_y$ are the exchange couplings while $h$ is the transverse field. This Hamiltonian can be fermionized through a Jordan-Wigner transformation \cite{mbengQuantumIsingChain2024} to obtain
\begin{equation}
     H = \sum_{j=1}^L \left[J c^\dagger_{j+1} c_j +  \eta c^\dagger_{j+1} c_j^\dagger + h.c. \right] + h \sum_{j=1}^L c_j^\dagger c_j,
\end{equation}
where $c_j$ and $c_j^\dagger$ are the usual fermionic annihilation and creation operators and where we have introduced the hopping amplitude $J=(J_x+J_y)/2$ and the pairing $\eta=(J_x-J_y)/2$. 
We consider to monitor the particle density on each lattice site, with a uniform rate $\gamma$, leading to the jump operator 
\begin{align}
L_j = c_j^\dagger c_j
\end{align} 
Since we have $L_j^\dagger L_j= c_j^\dagger c_j$ and given the form of the Eq.~\eqref{eq:sse_state} the state will stay Gaussian starting from a Gaussian initial one. In Fourier space $c_j = \frac{1}{\sqrt{2\pi}}\sum_k e^{-i~jk} c_k$ we can write the effective non-Hermitian Hamiltonian as a transverse field Ising chain in complex field~\cite{turkeshiEntanglementCorrelationSpreading2023} 
\begin{equation}
    H_\mathrm{nH} = \sum_k 2 J \cos(k) c_k^\dagger c_k  +  i\eta  \sin(k) \left[ c_k^\dagger c_{-k}^\dagger -   c_{-k} c_k  \right] - (i \frac{\gamma}{2} -h ) c_k^\dagger c_k .
    \label{eq:hamil_ising_kspace}
\end{equation}
We can briefly discuss some relevant limits of the monitored Ising chain, that we will discuss in more detail later on. For $\eta=0$ and $J\neq0$, the model reduces to free-fermions hopping with density monitoring, which has a strong U(1) symmetry associated to particle number conservation. The model has been studied at length both numerically and analytically~\cite{buchhold2022revealingmeasurementinduced,poboikoNonlinearSigmaModels2023,chahineEntanglementPhasesLocalization2024,krapivskyFreeFermionsLocalized2019}. For any $\eta\neq0$ the strong U(1) symmetry is broken to a strong discrete $Z_2$ symmetry. Numerically, the model has been shown to display a robust MIPT between a weak-monitoring phase with sub-volume scaling of the entanglement entropy and an area law phase~\cite{legalEntanglementDynamicsMonitored2024}. The transition can be crossed by increasing the monitoring rate $\gamma$ or the field $h$. We note how a similar entanglement transition is observed also in the no-click limit, which shares certain features of the phase diagram \cite{turkeshiEntanglementTransitionsStochastic2022, legalEntanglementDynamicsMonitored2024}. Finally, it is interesting to mention the case $J=0$ and $\eta\neq0$ in which the model has only next-neighbor pairing but no hopping. In this case one can show that by applying a particle-hole transformation the model can be mapped to a monitored Su-Schrieffer-Heeger model, which in the no-click limit
has PT-symmetry and a volume law phase~\cite{legal2023volume,soares2025symmetries}. Numerically, it was found that the effect of quantum jumps in this model is much stronger and the area-law is expected to be robust at large scales~\cite{legalEntanglementDynamicsMonitored2024}.

\subsection{The Keldysh Action for the Ising Chain }

\paragraph{Spaces conventions:}
In writing the Keldysh action for this problem we will have to introduce coherent fermionic fields carrying multiple labels corresponding to three different \emph{spaces}: the \emph{Keldysh} space for the  $+/-$ components on the Keldysh contour, the \emph{Nambu} space for pairing and anti-pairing terms present in the Hamiltonian of the fermionised Ising chain, and the \emph{Replica} space of size $R$, giving us in total $4R$ fields. Since the Keldysh and the Nambu spaces are of dimension $2$, we can parametrised them through Pauli matrices. In the following, we note $\sigma_\alpha$ the Pauli matrices in Keldysh space, with $\alpha \in \{0,x,y,z \}$  and $\tau_\beta$ the Pauli matrices in Nambu space, with $\beta \in \{0,x,y,z \}$ where the index $0$ refers to the $2 \times 2$ identity matrix. The Kronecker product is defined such that 
\begin{equation}
    A \otimes B = \begin{pmatrix}
        a_{1,1} B & a_{2,1} B \\
        a_{1,2} B & a_{2,2} B 
    \end{pmatrix},
\end{equation}
where $A$ is a $2 \times 2$ matrix. 
With these conventions, we first group the fermionic fields together into convenient spinor fields which we write in a $+/-$ Keldysh basis that we call $\mathcal{B}_{NK}$ and reads as
\begin{align}
    \Psi_{\mathrm{NK},r}^T &= \begin{pmatrix}
       \psi_{+,k,r} & \psi_{-,k,r}  & \overline{\psi}_{+,-k,r} & \overline{\psi}_{-,-k,r}
    \end{pmatrix}/ \sqrt{2}\\
   \overline{\Psi}_{\mathrm{NK},r} &= \begin{pmatrix}
        \overline{\psi}_{+,k,r} & \overline{\psi}_{-,k,r}  & \psi_{+,-k,r} & \psi_{-,-k,r}
     \end{pmatrix} / \sqrt{2}
\end{align}
Here $\mathrm{NK}$ indicates that we write the basis in the convention $\mathrm{Nambu} \otimes \mathrm{Keldysh}$, whereas $\mathrm{KN}$ would mean that we consider $\mathrm{Keldysh} \otimes \mathrm{Nambu}$. The index $r$ denotes the replica index, while $k,-k$ are momentum states coupled by pairing/antipairing terms in the Hamiltonian, resulting in a non-trivial Nambu structure. In the following $\Psi_{\mathrm{NK}}$ should be understood as $\bigoplus_{r=1}^R \Psi_{\mathrm{NK},r}$. It is also convenient to consider the Larkin-Ovchinikov (LO) basis defined as $\psi_{1,2} = (\psi_+  \pm  \psi_- )/\sqrt{2}$ and $\overline{\psi}_{1,2} = (\overline{\psi}_+ \mp \overline{\psi}_- )/\sqrt{2}$, since this basis reveals the causality structure and presents the usual structure in terms of retarded and advanced Green's functions (See Appendix \ref{app:causality_struct}). We call this basis $\mathcal{B}^{(LO)}_{KN}$ and write
\begin{align}
    \Phi_{\mathrm{KN},r}^T &= \begin{pmatrix}
        \psi_{1,k,r} & \overline{\psi}_{2,-k,r} & \psi_{2,k,r} & \overline{\psi}_{1,-k,r}
    \end{pmatrix} / \sqrt{2}\\   \overline{\Phi}_{\mathrm{KN},r} &=\begin{pmatrix}
        \overline{\psi}_{1,k,r}    & \psi_{2,-k,r} & \overline{\psi}_{2,k,r}& \psi_{1,-k,r}
    \end{pmatrix} / \sqrt{2},
    \label{eq:basis_phi_kn}
\end{align}
where $\Phi$ indicates the use of the LO basis and $\mathrm{KN}$ the use of the other ordering convention. We note that 
\begin{equation}
    \overline{\Phi}_{KN, r}^T(k)= C \Phi_{KN, r}(-k), \quad \mathrm{with}\quad C = \sigma_x \otimes \tau_x 
    \label{eq:spinor_constraint}
\end{equation}
and in real space this simply reads as $\overline{\Phi}_{KN, r}^T= C \Phi_{KN, r}  $. Using the Keldysh action Eq.~\eqref{eq:action_gen_eff} and the Fourier transform of the effective Hamiltonian Eq.~\eqref{eq:hamil_ising_kspace}, we can write the quadratic part of the action for the Ising chain in the $\mathcal{B}_{NK}$ basis as
\begin{align}
    S_\mathrm{nH} &= \int_{t_i}^{t_f} dt \int_{-\pi}^{\pi} \frac{dk}{2\pi}\sum_{r=1}^R ~ \overline{\Psi}_{NK,r} \left[ G_{0,\pm,NK}^{-1} \right]   \Psi_{NK,r}, \\
     G_{0,\pm,NK}^{-1} &= i\partial_t (\tau_0 \otimes \sigma_z)  - \left[2J\cos(k) -h\right] (\tau_z \otimes \sigma_z ) - 2 i \eta  \sin(k) ( i\tau_y \otimes \sigma_z) + i \frac{\gamma}{2} ( \tau_z \otimes \sigma_0).
\end{align}
where $G_{0,\pm, NK}^{-1}$ is the inverse fermionic propagator associated to the non-Hermitian effective Hamiltonian in Eq.~\eqref{eq:hamil_ising_kspace}.
Similarly, by finding the matrix to pass from $\Psi_{NK}$ to $\Phi_{KN}$, we can rewrite the action in the $\mathcal{B}^{(LO)}_{KN}$ basis as \footnote{In general the matrix to pass from $\Psi_{NK}$ to $\Phi_{KN}$ and from $\overline{\Psi}_{NK}$ to $\overline{\Phi}_{KN}$ are not complex conjugates from each other.} $S_\mathrm{nH} = \int_{-\infty}^{\infty}  \frac{d\omega}{2 \pi} \int_{-\pi}^{\pi} \frac{dk}{2\pi}\sum_{r=1}^R ~ \overline{\Phi}_{KN,r} \left[ G_{0,KN}^{-1} \right]   \Phi_{KN,r}$, where we consider $t_i \rightarrow - \infty$ and $t_f \rightarrow \infty$, and perform a temporal Fourier transform, effectively implementing $i\partial_t \leftrightarrow \omega$ and with 
\begin{align}
     G_{0,KN}^{-1} &= g_{0,KN}^{-1}+  i \frac{\gamma}{2} (\sigma_x \otimes \tau_z ),
    \label{eq:def_bare_green_nH} \\
     g_{0,KN}^{-1} &= \omega(\sigma_0 \otimes \tau_0) - \left[2J \cos(k) -h \right] (\sigma_0 \otimes \tau_z) + 2 \eta \sin(k) (\sigma_0 \otimes \tau_y).
     \label{eq:def_bare_green_symm}
\end{align}
The matrix $g_{0,KN}^{-1}$ is the inverse fermionic propagator associated to the Hermitian part of the effective Hamiltonian, and reads in explicit form
\begin{equation}
    g_{0,KN}^{-1}  = \sigma_0 \otimes  \begin{pmatrix}
        \omega - 2J \cos(k) + h & 2 i \eta \sin(k)  \\
       - 2 i \eta \sin(k) &  \omega + 2J \cos(k) - h 
    \end{pmatrix}.
\end{equation}
We note that as compared to a conventional fermionic Keldysh action, in which causality imposes a specific structure to the action leading to vanishing of certain correlators due to causality~\cite{thompsonFieldTheoryManybody2023}, here we have non-zero terms due to the measurement back action $\gamma$ which enters in the non-Hermitian Hamiltonian, leading to the structure $\sigma_x \otimes \tau_z $ in Keldysh-Nambu space in Eq.~(\ref{eq:def_bare_green_nH}).

Moving to the quantum jumps part of the action, due to the density monitoring with jump operator $L_x = c_x^\dagger c_x$, we can write it as  
\begin{equation}
    S_\mathrm{QJ} = -i\gamma \int dt  dx ~   \prod_{r=1}^R \overline{\psi}_{x,+,r}(t) \psi_{x,+,r}(t) \overline{\psi}_{x,-,r}(t)\psi_{x,-,r}(t) = -i\gamma \int dt dx~ \prod_{r=1}^R V_{r}(x,t), 
\end{equation}
and in the Larkin-Ovchinikov basis we get 
\begin{equation}
    V_{r}(x,t)    = \overline{\psi}_{x,r,1}(t) \psi_{x,r,2}(t) \overline{\psi}_{x,r,2}(t) \psi_{x,r,1} (t).
\end{equation} 
This term is local in space and time, but Keldysh Green's function 
are discontinuous at equal time. To avoid this issue, we must regularise the Green function to make it continuous and still preserve the same action \cite{yangKeldyshNonlinearSigma2023, poboikoNonlinearSigmaModels2023,starchlGeneralizedZenoEffect2024}. This process is detailed in Appendix \ref{app:regularisation_green_function}, and leads to the following regularised interaction
\begin{equation}
    V^\mathrm{reg}_{r}(x,t) = \frac{1}{4} + \frac{1}{2}\left[ \overline{\psi}_{x,r,1}(t) \psi_{x,r,2}(t)+  \overline{\psi}_{x,r,2} (t)\psi_{x,r,1}(t) \right] - \overline{\psi}_{x,r,1}(t)\psi_{x,r,1}(t)\overline{\psi}_{x,r,2}(t)\psi_{x,r,2}(t),
\end{equation}
as previously found in Ref. \cite{poboikoNonlinearSigmaModels2023,starchlGeneralizedZenoEffect2024}.  This regularised vertex can be conveniently rewritten as
\begin{equation}
    V_{r}^\mathrm{reg}(x,t)  =  \frac{1}{4} \exp\left[ 2 ~\overline{\Phi}_{\mathrm{KN},r} (\sigma_x \otimes \tau_z ) \Phi_{\mathrm{KN},r} \right] \quad \mathrm{and} \quad  S_\mathrm{QJ} = -i\gamma \int dt \sum_{x=1}^L V_{R}^\mathrm{reg}(x,t),
\end{equation} 
with $V_{R}^\mathrm{reg}(x,t) = \prod_{r=1}^R V_{r}^\mathrm{reg}(x,t) $, where contrary to  Ref. \cite{poboikoNonlinearSigmaModels2023,starchlGeneralizedZenoEffect2024}, we have an extra Pauli matrix due to the Nambu Structure. 

All in all, we obtain the following replicated Keldysh action for the monitored Ising chain,
\begin{equation}
    Z = \int \mathcal{D}[\Phi_{\mathrm{KN}}] \exp\left[ i \int dt \int_{-\pi}^{\pi} \frac{dk}{2\pi}  ~ \overline{\Phi}_{\mathrm{KN}} [G_{0,\mathrm{KN}}^{-1} \otimes \mathbbm{1}_R ] \Phi_{\mathrm{KN}}  + \gamma \int dt \sum_{x=1}^L   V_{R}(x,t) \right].
 \end{equation}
The quadratic part of the action accounts for unitary evolution and measurement back-action, while the interactions are due to quantum jumps. In the following sections, we will present an analysis of this action using techniques developed for disordered systems and Anderson localization, leading to an effective theory described by a Non-Linear Sigma Model~\cite{eversAndersonTransitions2008}.

\section{Non-Linear Sigma Model for the Monitored Ising Chain}\label{sec:NLSM_Ising}

\subsection{Derivation of the Bosonic Action}

We now aim to replace the Grassmann integration over fermionic coherent fields with an integration over bosonic collective fields which can be achieved via a Hubbard-Stratonovich transformation involving the auxiliary variables $\mathcal{G}$ and $\Sigma$, both $4R \times 4R$ matrices and function of space-time\footnote{In the following, we will sometime write the space time dependence as extra matrix dimension, especially when using traces}. We are thus starting from 
\begin{equation}
    Z = \int \mathcal{D}[\Phi_{\mathrm{KN}}] \exp\left[ i \int dt \int_{-\pi}^{\pi} \frac{dk}{2\pi}  ~ \overline{\Phi}_{\mathrm{KN}} [G_{0,\mathrm{KN}}^{-1} \otimes \mathbbm{1}_R ] \Phi_{\mathrm{KN}}  + \gamma \int dt dx ~    V_{R}(x,t) \right],
 \end{equation}
and introduce the Dirac function $\delta\left( \mathcal{G}(x,t) + i \Phi_\mathrm{KN}(x,t) \overline{\Phi}_\mathrm{KN}(x,t) \right)$, with $\mathcal{G}(x,t)$ a bosonic Hermitian field local in space an time. By using the identity resolution $ \int \mathcal{D}\mathcal{G}~\delta\left( \mathcal{G} + i \Phi_\mathrm{KN} \overline{\Phi}_\mathrm{KN} \right) =1 $, we enforce the relation $\mathcal{G}(x,t) = -i \Phi_\mathrm{KN}(x,t) \overline{\Phi}_\mathrm{KN}(x,t)$. 
As explained in \cite{poboikoNonlinearSigmaModels2023}, we can decouple the interaction vertex by effectively performing the average of the interaction vertex with respect to the Gaussian action $\overline{\Phi}_{KN} \mathcal{G}^{-1} \Phi_{KN} $. We obtain the following decoupled vertex\footnote{We are writing $\sqrt{\mathrm{det}\left[ \circ \right] }$ insead of $\mathrm{Pf}\left[\circ \right]$ for convenience, but we should be careful when doing so since this introduce an ill defined sign, that is here verified by computing the Pfaffian, see the Appendix \ref{app:symplectic} for details }
\begin{align}
    \nonumber V^\mathrm{dec}[\mathcal{G} ](x,t) &= \int \frac{\mathcal{D}\Phi_{\mathrm{KN}} }{\mathrm{Pf}\left[-i\mathcal{G}^{-1}\right]} \exp\left[i  \frac{1}{2}\overline{\Phi}_{\mathrm{KN}}  \mathcal{G}^{-1} \Phi_{\mathrm{KN}} \right] V_{R}^\mathrm{reg}(x,t) \\
    &  = \frac{1}{4^R} \sqrt{ \mathrm{det}\left[ \mathbbm{1}_{4R} - i 4 \mathcal{G}(x,t) (\sigma_x \otimes \tau_z \otimes \mathbbm{1}_R ) \right]  },
\end{align}
where unlike the projective measurement case in a tight binding chain studied in \cite{poboikoNonlinearSigmaModels2023,poboikoMeasurementInducedPhaseTransition2024}, we here have a single term (since jumps do not have several outcomes). The square root is due to the Nambu basis, which imposes that $\overline{\Phi}_{\mathrm{KN}}$ is not independent from $\Phi_{\mathrm{KN}} $ anymore, details about this symplectic structure can be found in Appendix \ref{app:symplectic}. This procedure leads to the following action
\begin{align}
    Z & = \int \mathcal{D} \mathcal{G}\mathcal{D}[\Phi_{\mathrm{KN}}] ~ \delta\left( \mathcal{G} + i \Phi_\mathrm{KN} \overline{\Phi}_\mathrm{KN} \right)\exp\left( i S\left[ \mathcal{G}, \Phi_{\mathrm{KN}}\right]  \right)  \\   
    S &= \int dt \int_{-\pi}^{\pi} \frac{dk}{2\pi}  ~ \overline{\Phi}_{\mathrm{KN}} [G_{0,\mathrm{KN}}^{-1} \otimes \mathbbm{1}_R ] \Phi_{\mathrm{KN}}  -i  \gamma \int dt dx~ V^\mathrm{dec}[\mathcal{G}(x,t)].
 \end{align}
To obtain a regular action, we use the following representation of the Dirac function 
\begin{equation}
    \delta\left( \mathcal{G} + i \Phi_\mathrm{KN} \overline{\Phi}_\mathrm{KN} \right) = \lim_{\epsilon \rightarrow 0 } \int \mathcal{D} \Sigma ~\exp \left[- \frac{1}{2 \epsilon} \mathrm{Tr}\left( \mathcal{G} + i \Phi_\mathrm{KN} \overline{\Phi}_\mathrm{KN} \right)^2 - \frac{\epsilon}{2} \mathrm{Tr}\left( \Sigma^2 \right) \right],
\end{equation}
using another Hermitian field $\Sigma(x,t)$, also local in space and time. By performing the change of variable $\Sigma \rightarrow \Sigma + \tfrac{i}{\epsilon}(\mathcal{G} + i \Phi_\mathrm{KN} \overline{\Phi}_\mathrm{KN} )$, one gets (omitting the superscript ‘dec’)\footnote{We use in the computation the property of the Dirac function, even if the limit $\epsilon \rightarrow 0$ is taken after. Moreover, the change of variables eliminates the limit explicitly by incorporating it into the bounds of the integral.  }
\begin{align}
    Z &= \int \mathcal{D} \mathcal{G}\mathcal{D}  \Sigma \mathcal{D}[\Phi_\mathrm{KN}] ~ \exp\left[ i S \right],  \\
    S &= \int dt \int_{-\pi}^{\pi} \frac{dk}{2\pi}  ~ \overline{\Phi}_{\mathrm{KN}} \left[G_{0,\mathrm{KN}}^{-1} \otimes \mathbbm{1}_R  +i \Sigma  \right] \Phi_{\mathrm{KN}}  - \mathrm{Tr}\left( \Sigma \mathcal{G} \right) - i \gamma \int dt dx~ V[\mathcal{G}(x,t) ].
 \end{align}
 We can finally perform the Grassmann integral, to integrate out the fermionic fields $\Phi_\mathrm{KN}$, and obtain\footnote{We note here that in Fourier space the two spinors are independent since they refer to different $k$ modes, thus the Grassmann integrals gives the usual determinant. It is also the reason of the absence of the prefactor $1/2$ }
 \begin{multline} 
         S =   -i\mathrm{Tr}\left\{ \ln\left[ 2( -iG_{0,\mathrm{KN}}^{-1} \otimes \mathbbm{1}_R + \Sigma ) \right] \right\}    - \mathrm{Tr}\left( \Sigma \mathcal{G} \right)  \\ - i \gamma \int dt dx~  \frac{1}{4^R} \sqrt{\mathrm{det}\left[ 1 - 4i \mathcal{G}(x,t) (\sigma_x \otimes \tau_z \otimes \mathbbm{1}_R) \right]} .  
    \label{eq:full_action}
 \end{multline}
This action is a functional of slowly varying collective fields $\mathcal{G}(x,t),~\Sigma(x,t)$ which are $4R\times4R$ matrices in replica, Nambu and Keldysh space, and slow function of space and time. It is worth commenting on the relation between this result and previous literature on free-fermions with U(1) symmetry. For projective measurements the two projectors entering the Kraus operators lead to an action with two determinant contributions and no structure in Nambu space~\cite{poboikoNonlinearSigmaModels2023}, except for particle-hole symmetry~\cite{poboikoMeasurementinducedTransitionsInteracting2025}. In our case, instead, one of the two Kraus operator is close to the identity and taken into account exactly in the non-Hermitian Hamiltonian, which leaves us with only a single determinant in Eq.~(\ref{eq:full_action}). We note that with respect to Ref.~\cite{starchlGeneralizedZenoEffect2024} we take into account exactly the non-Hermiticity of the action, without needing to add a counter-term.

In the remaining of this work, we will study the saddle point of the action in Eq.~(\ref{eq:full_action}) and then expand in non-linear fluctuations around the saddle point to obtain an effective field theory, which will take the form of a Non-Linear Sigma Model.

\subsection{The Saddle Point and its Solution }
\label{sec:ising_saddle}

We now solve the effective action in Eq.~\eqref{eq:full_action} within a saddle point method, which is expected to be controlled in the limit of weak measurements $\gamma/J\ll 1$\cite{poboikoNonlinearSigmaModels2023}. We begin by seeking a replica-symmetric saddle point and will later consider fluctuations that break this symmetry. Under this assumption, we can equivalently consider $R=1$ to simplify the computation, we note that we should retrieve the unconditional dynamics described by a Lindblad master equation.

It is easier to work with the interaction vertex in real space, we thus temporarily transform back the correlation matrix in real space. In real space, the spinors $\overline{\Phi}_\mathrm{KN}$ and $\Phi_\mathrm{KN}$ are not independent from each other, which introduces an extra factor $1/2$ in the integration of the fermionic fields, leading to 
\begin{equation}
    S = -i \mathrm{Tr}\left[\frac{1}{2}\ln\left(2\left[ -i G_{0,\mathrm{KN}}^{-1} + \Sigma \right]\right) \right] -\mathrm{Tr}\left[\Sigma \mathcal{G}\right] -  i \gamma \int dt dx ~ \frac{1}{4} \sqrt{\mathrm{det}[\mathbbm{1}_4 - 4 i \mathcal{G}(x,t) ( \sigma_x \otimes \tau_z) ]}.
    \label{eq:action_full_KN_real}
\end{equation}
Since $\mathcal{G}(x,t) \propto -i \Phi_\mathrm{KN}(x,t)\overline{\Phi}_\mathrm{KN}(x,t)$ and Eq.~\eqref{eq:spinor_constraint}, the matrix  $\mathcal{G}$ is $4\times 4$ symplectic matrix ($\mathcal{G} = - C \mathcal{G}^T C  $), and we can use the relation 
\begin{equation}
    \frac{1}{4}\sqrt{\mathrm{det}[\mathbbm{1}_4 - 2i \mathcal{G} (\sigma_x \otimes \tau_z )]} =   4\sqrt{\mathrm{det}[\mathcal{G}] } - \frac{i}{2}\mathrm{Tr} \left[\mathcal{G}~(\sigma_x \otimes \tau_z ) \right] +  \frac{1}{4}. 
\end{equation}
This relation due to the symplectic symmetry is explained in Appendix \ref{app:symplectic}.  By considering the variation of the action $S$ with respect to $\mathcal{G}$ and $\Sigma$, and set it to zero, we obtain the saddle point equations
\begin{align}
     \mathcal{G}& = -i \frac{1}{2}\left( -i G_{0,\mathrm{KN}}^{-1} + \Sigma \right)^{-1} \qquad \mathrm{for} \qquad \frac{\delta S }{\delta \Sigma} = 0,  \label{eq:saddlepoint_eq_1}  \\ 
     \Sigma & =  - 2 i\gamma   \sqrt{\mathrm{det}[\mathcal{G}]} \mathcal{G}^{-1} - \frac{\gamma}{2}  (\sigma_x \otimes \tau_z)  \qquad \mathrm{for} \qquad \frac{\delta S }{\delta \mathcal{G}} = 0, \label{eq:saddlepoint_eq_2}  
\end{align}
where $\delta S / \delta \Sigma$ stands for $\left(\delta S / \delta \Sigma \right)_{ij} = \partial S / \partial \Sigma_{ij}$ with $i,j$ generic indices\footnote{
We use Jacobi's formula reading as 
\begin{equation}
     \frac{d}{dA} \mathrm{Pf}\left[GA\right] = \frac{1}{2}  A^{-1}  \mathrm{Pf}\left[GA\right],
\end{equation}
for generic matrices $G$ and $A$.  
 }. Because of the form of Eq.~\eqref{eq:saddlepoint_eq_1}, we propose the following ansatz $\mathcal{G} = -i  Q/4 $, where $Q$ is a $4R\times 4R$ matrix dependent on space and time, which satisfies $\mathrm{det}\left[Q \right]=1 $\footnote{This is different from the $U(1)$ case \cite{poboikoNonlinearSigmaModels2023}: since we have $4$ eigenvalues, that will be $+1$ and $-1$ the determinant will be $1$. In this context, the objective is to find a single valid saddle point, hence, it is acceptable to impose arbitrary constraints as long as they are internally consistent.}.
 Hence, by using this ansatz in Eq.~\eqref{eq:saddlepoint_eq_2} we get the following parametrisation
\begin{align}
    \Sigma &=   -2i \gamma \frac{4}{4^2} \sqrt{( -i )^4 \mathrm{det}[Q]}( -i Q)^{-1} - \frac{\gamma}{2} ( \sigma_x \otimes \tau_z) =     \frac{\gamma}{2} \left[ Q -  ( \sigma_x \otimes \tau_z )  \right] .
\end{align} 
We obtain the following ansatz 
\begin{equation}
    \mathcal{G} =  -i  \frac{Q}{4}, \qquad \Sigma = \frac{\gamma}{2} \left[     Q -  ( \sigma_x \otimes \tau_z )  \right],
    \label{eq:ansatz_saddle_point}
\end{equation}
that we will now use to solve the saddle point equation Eq.~\eqref{eq:saddlepoint_eq_1}.  \\

By using the ansatz of Eq.~\eqref{eq:ansatz_saddle_point} in the saddle point equation Eq.~\eqref{eq:saddlepoint_eq_1}, we notice that $-i G_{0,\mathrm{KN}}^{-1} + \Sigma = -i g_{0,\mathrm{KN}}^{-1} + \tfrac{\gamma}{2}Q$. In other words, the term $-\frac{\gamma}{2} \sigma_x \otimes \tau_z$ appearing in Eq.~(\ref{eq:ansatz_saddle_point}) for $\Sigma$ exactly cancels with the term entering the non-interacting Green's function $-i G_{0,\mathrm{KN}}^{-1}$, leading to 
\begin{equation}
    \frac{Q}{2} =  \left( -i g_{0,KN}^{-1}+ \frac{\gamma}{2}Q   \right)^{-1}.
\end{equation}
Therefore, we recover a saddle point equation which is consistent with expectations: in the replica-symmetric case ($R=1$), we should reproduce the Lindbladian result and, in particular, restore the causal structure that was previously disrupted by the non-Hermitian term in the non-interacting action. Having this causal structure (preserved in the inversion of the matrix), we can write the equation in the RAK basis 

\begin{equation}
    \frac{Q}{2} =\frac{1}{2} \begin{pmatrix} \Lambda^R & \Lambda^K \\
    0 & \Lambda^A \end{pmatrix}=  \left[  -i \begin{pmatrix} g_{0,KN}^{-1,R} & 0 \\
        0 & g_{0,KN}^{-1,A} \end{pmatrix} +  \frac{\gamma}{2} \begin{pmatrix} \Lambda^R & \Lambda^K \\
            0 & \Lambda^A \end{pmatrix}  \right]^{-1},
\end{equation}
where $R$ (resp. $A$) subscript denotes the retarded (resp. advanced) block of the matrix. To go further and solve the equation, we need to diagonalize $g_{0,KN}^{-1,R}$ and assume that we can diagonalize in the same basis $Q^R$ (like usually done in BCS theory \cite{yurkevichNonlinearSigmaModel2001,feigelmanKeldyshActionDisordered2000}). The matrix $g_{0,KN}^{-1,R}$ has for eigenvalues 
\begin{equation}
    \mu_{\pm} = \omega \pm \sqrt{4 \eta^2 \sin(k)^2 + (h - 2 J \cos(k))^2 } =  \omega \pm \xi_k.
\end{equation}
where $\xi_k$ is the quasiparticle dispersion relation of the transverse field Ising chain. By going to the diagonalisation basis, we obtain the following self consistent scalar equation 
  
\begin{equation}
   \frac{\mathcal{\lambda}^R}{2} = i \int_{-\infty}^{\infty} ~\frac{d\omega}{2\pi} \int_{-\pi}^{2\pi} ~ \frac{dk}{\pi} \frac{1}{ \omega \pm \xi_k + i \gamma \lambda^R/2 } =   \frac{1}{2} \mathrm{Sgn}[\lambda^R],
\end{equation}
which implies that  $\lambda^R = \pm1$\footnote{We note that contrary to the usual case of \cite{kamenevFieldTheoryNonEquilibrium2011}, here since the interaction term is local in space and time, the decoupling only generate terms $\mathcal{G}(x,t)$ and the saddle point is thus also local in spacetime. This explains why one both has an integral on frequency $\omega$ and wave vector $k$} (details about this computation can be found in Appendix \ref{app:saddle_point_int}). To preserve the right causality, we must have $\lambda^R = 1$, and by proceeding similarly for the advanced component we get $\lambda^A = -1 $. The saddle point $\Lambda$ then presents the usual structure 
\begin{equation}
    \Lambda = \begin{pmatrix}
        \mathbbm{1}_{2} & \mathcal{F} \\
        0 & -\mathbbm{1}_{2}
    \end{pmatrix}_{KN}.
\end{equation}
The Keldysh component $\Lambda^K$ is not fixed by the stationary point equation but can be parametrised by the distribution function $\mathcal{F}$ such that $\Lambda^K = \Lambda^R \mathcal{F} -  \mathcal{F} \Lambda^A$ where this function can be determined from the boundary conditions \cite{kamenevFieldTheoryNonEquilibrium2011,poboikoNonlinearSigmaModels2023}. Finally, the saddle point correlation function, reads as $\mathcal{G}^{-1,R} = 2i \left( -i g_{0,KN}^{-1,R} + \gamma Q^R /2 \right)$ for the retarded components, and can be explicitly written 
\begin{equation}
    \mathcal{G}^{-1,R}(k,\omega) = 2 \begin{pmatrix}
        \omega - 2J\cos(k)+h + i\gamma/2 & -2 i \eta  \sin(k)\\ 2 i \eta \sin(k) &  \omega + 2J\cos(k) - h + i\gamma/2
    \end{pmatrix}.
\end{equation}
As a sanity check, we can compare this result to the one directly derived from the Lindblad master equation and find perfect agreement, see Appendix~\ref{app:lindblad}.

The saddle point for the replica symmetric sector (or for $R=1$ as done in this section) describes the physics of the average state, which evolves towards infinite temperature due to dephasing arising from averaging over the measurement outcomes. Interestingly the saddle point method for $R=1$ turns out to be equivalent to a diagrammatic resummation of the interaction vertex via self-consistent Born approximation, which is exact for the Lindbladian problem. However, for $R\neq 1$ this is no longer the case and one needs to include higher order corrections, as we will do by means of the NLSM. Indeed, we obtain here a saddle point that satisfies $Q^2 =1$, which is characteristic of the non-linear constraint in the  NLSM. As we will see, fluctuations that violate this constraint will become massive, since in that case  $\mathrm{Tr}\left[ \Sigma \mathcal{G}\right]$ contributes significantly. 

\subsection{The Symmetries of the Replicated Action }
\label{sec:symmetries}

The goal of this section is to determine the symmetries of the replicated Keldysh action. This step is crucial to derive the effective field theory of our model, which will be described in terms of slow-modes corresponding to Goldstone modes associated to the breaking of these symmetries. 

Finding the symmetries of our problem can be viewed as finding the rotation $\mathcal{R}$ such that $\Phi \rightarrow \mathcal{R} \Phi$ and $\overline{\Phi}_\mathrm{KN} \rightarrow \overline{\Phi}_\mathrm{KN} \mathcal{R} $ with $\mathcal{R} \in U(4R)$, which are leaving the action invariant. Since $\mathcal{G}(x,t) \propto -i \Phi_\mathrm{KN}(x,t)\overline{\Phi}_\mathrm{KN}(x,t)$, this constraint reads at the level of $\mathcal{G}$ as $\mathcal{G} = \mathcal{R} \mathcal{G} \mathcal{R}^{-1}$. 

This will restrict the symmetry group of our field theory to a subgroup $G$ of the full $U(4R)$, which we want to identify. Furthermore, to parametrize the fluctuations around the saddle point, and obtain the effective action, we want to identify the symmetry operations that leave the saddle point invariant, which we will identify with the group $H$. Ultimately, this will allow us to obtain the manifold of the NLSM as the quotient group $G/H$ and the associated class in the Altland-Zirbauer ten-fold classification of disordered systems \cite{ludwigTopologicalPhasesClassification2016, eversAndersonTransitions2008}. The full NLSM will be then derived in the next section.



The first step is to determine the constraints that the rotation matrix must satisfy in order to leave the action of Eq.~\eqref{eq:full_action} invariant. First, since $\mathcal{R}$ is a rotation, it must fulfil $\mathcal{R}^\dagger = \mathcal{R}^{-1}$. Then as we have seen previously in Eq.~\eqref{eq:spinor_constraint}, the fermionic field  and its conjugate are not independent, imposing the constraint $C \mathcal{G} C = - \mathcal{G}^T  $. Since, ultimately, we will parametrize the entire solution manifold using $Q$, viewed as a non-trivial rotation of the saddle point $\Lambda$, the rotations that define this manifold must preserve $Q$, leading to the condition $Q = \mathcal{R} Q \mathcal{R}^{-1}$. Hence, we find $\mathcal{R} Q \mathcal{R}^{-1}  = \left[ \mathcal{C} (\mathcal{R}^{-1} )^T \mathcal{C} \right] Q \left[ \mathcal{C}  \mathcal{R}^T \mathcal{C} \right]$ that reduces to the constraint $\mathcal{C}  \mathcal{R}^* \mathcal{C}  = \mathcal{R}$, leading to the two structural constraints
\begin{equation}
    \nonumber \mathcal{R}^\dagger = \mathcal{R}^{-1} \quad (\boldsymbol{C0})\quad \mathrm{and} \quad  \mathcal{C}  \mathcal{R}^* \mathcal{C}  = \mathcal{R} \quad (\boldsymbol{C1}). 
\end{equation}
We then want the rotations to leave invariant the action in Eq.~\eqref{eq:full_action}, meaning that it should commute with the interaction term due to quantum jumps, whose structure is $\mathbbm{1}_{R}\otimes \sigma_x \otimes \tau_z$, and the bare Green function $G_{0,KN}^{-1}$ defined in Eq.~\eqref{eq:def_bare_green_nH}, and imposing  
\begin{equation}   \begin{cases}
    (\boldsymbol{C2}) : \quad \left[ \mathcal{R} ,\mathbbm{1}_R \otimes \sigma_x \otimes \tau_z  \right] = 0 & (\mathrm{Interaction ~ term})\\
    (\boldsymbol{C3}) : \quad   \left[ \mathcal{R} , \mathbbm{1}_R \otimes\sigma_0 \otimes \tau_z \right]  = 0 & (\mathrm{Tight~ Binding ~ term}) \\
    (\boldsymbol{C4}) : \quad    \left[ \mathcal{R} ,\mathbbm{1}_R \otimes  \sigma_0 \otimes \tau_y \right]  = 0 & (\mathrm{Pairing ~ term}) .
    \end{cases} \label{eq:constraints_C2_C4}
\end{equation}
It is worth mentioning that the time-derivative term does not contribute, as it is proportional to the identity (which was a motivation for this basis choice). Furthermore, the non-Hermiticity 
of $G_{0,KN}^{-1}$ does not play a distinct role in determining the symmetries, as it imposes the same constraint as the quantum jumps term. To characterize the symmetries of our theory, we present two methods to proceed. The first one counts the remaining number of degrees of freedom once we implement the different constraints of the action, whereas the second one directly builds the rotation matrix. 

\subsubsection{Method 1 : Counting Free Parameters }

In this method we will parametrize the rotation matrix and count the free parameters remaining after imposing the constraints. This particular number of free parameters usually allow us to retrieve the corresponding symmetry group easily.

We parametrize the $4R \times 4R$ rotation matrix $\mathcal{R}$ as
\begin{equation}
    \mathcal{R} = \exp \left[ X\right] = \exp \left[ \sum_{i,j} W_{ij} \left(\sigma_i \otimes \tau_j  \right) \right],
\end{equation}
where $W_{ij}$ are $R \times R$ skew-symmetric matrices ($W_{ij}^\dagger = - W_{ij}$) and the indices $i,j \in \{0,x,y,z\}$ refer to the Pauli indices. With this parametrisation, we have a generic unitary matrix, and can deduce the restrictions the constraints are enforcing. The constraints are of the form $U \exp \left[ X \right] U^\dagger = \exp \left[ X \right]$, and we want to be able to use the injectivity of the exponential to be able to deduce that this constraint imply $UXU^\dagger = X $. This is wrong in general, but in our case since we are looking for continuous symmetries, we can consider the $W_{ij}$ to be small\footnote{As long as the spectrum of $X$ belongs to $\left]-i\pi , i\pi\right]$, we cannot have eigenvalues separated by an integer number of times $2\pi$, and thus the exponential is injective. } which guarantee injectivity. \\  


\begin{table}[t]
    \centering
    \begin{tikzpicture}
        \node (array1) {
            $\begin{tabular}{||c|| c c c c||}
                \hline
                & $\sigma_0$ & $\sigma_x$ & $\sigma_y$ & $\sigma_z$ \\ [0.5ex]
                \hline\hline
                $\tau_0$ & A,R &  A,R &  A,R & S,I \\
                \hline
                $\tau_x$ & A,R & A,R & A,R &  S,I \\
                \hline
                $\tau_y$ & A,R  & A,R&  A,R & S,I  \\
                \hline
                $\tau_z$ & S,I  & S,I  & S,I & S,I \\
                \hline
            \end{tabular}  $
        }; 
    \end{tikzpicture}
    \caption{Structure of the $W_{ij}$ matrices under only the structural constraints ie. the skew-symmetry imposed by the unitarity and $(\boldsymbol{C1})$. The entry $\sigma_i \otimes \tau_j$ correspond to the matrix $W_{ij}$. } 
    \label{tab:struct}
\end{table}

To proceed we now check what restriction on each $W_{ij}$, constraints are imposing. Lets first consider a specific case with $W_{zx}$. From $(\boldsymbol{C1})$, since $\left\{ C, \sigma_x \otimes \tau_z \right\} = 0 $, we obtain $W_{zx}^* = - W_{zx}$. Then because $[\sigma_x \otimes \tau_z, \sigma_x \otimes \tau_z ]= 0 $ and $[\sigma_0 \otimes \tau_z, \sigma_x \otimes \tau_z ]= 0 $, the constraints $(\boldsymbol{C2})$ and $(\boldsymbol{C3})$ give no further restrictions. If we do not enforce $(\boldsymbol{C4})$, we thus obtain that $W_{zx}$ is symmetric and purely imaginary. On the other hand, if $\eta \neq 0$, the constraint $(\boldsymbol{C4})$ gives $W_{zx} = - W_{zx}$ since $\{\sigma_0 \otimes \tau_y, \sigma_x \otimes \tau_z\}=0$, and we obtain $W_{zx}=0$. By doing similar computation for all the matrices $W_{ij}$ we can dress tables where: we indicate by a blank if $W_{ij}=0$ and otherwise precise if $W_{ij}$ is antisymmetric (A) or symmetric (S) and if it is real (R) or imaginary (I). We first write the array which corresponds to only enforcing $(\boldsymbol{C1})$ in Table~\ref{tab:struct}, before discussing the other different situations. 

Depending on the surviving entries in the table, the number of independent parameters changes, resulting in different symmetry groups.

\subsubsection{The $U(1)$ Limit of the Ising Chain }

\begin{table}[t]
    \centering
    \begin{tikzpicture}
        \node (array1) {
            $\begin{tabular}{||c|| c c c c||}
                \hline
                & $\sigma_0$ & $\sigma_x$ & $\sigma_y$ & $\sigma_z$ \\ [0.5ex]
                \hline\hline
                $\tau_0$ & A,R & A,R  &   & \\
                \hline
                $\tau_x$ &  &  &   &   \\
                \hline
                $\tau_y$ &   &  &   &   \\
                \hline
                $\tau_z$ &  S,I  & S,I   &  &  \\
                \hline
            \end{tabular}  $
        };
        \node[left=0.3cm of array1, yshift=1cm] {a)};

        \node (array2) at (7.5cm,0) {
            $\begin{tabular}{||c|| c c c c||}
                \hline
                & $\sigma_0$ & $\sigma_x$ & $\sigma_y$ & $\sigma_z$ \\ [0.5ex]
                \hline\hline
                $\tau_0$ & A,R &   &   &  \\
                \hline
                $\tau_x$ &  &  &   &  \\
                \hline
                $\tau_y$ &   &  &   &   \\
                \hline
                $\tau_z$ & S,I  &   &  &  \\
                \hline
            \end{tabular} $
        };
        \node[left=0.3cm of array2, yshift=1cm] {b)};
    \end{tikzpicture}
    \caption{Case of the U(1) limit of the Ising chain : $\eta = 0$ and $J\neq 0$, $\gamma\neq 0$ - a) Structure of the $W_{ij}$ matrice under constraints $(\boldsymbol{C0})$, $(\boldsymbol{C1})$, $(\boldsymbol{C2})$ and $(\boldsymbol{C3})$, b) Structure of the $W_{ij}$ under the constraints of a) and saddle point invariance additionally. }
    \label{tab:no_pairing}
\end{table}

We first consider the case without pairing terms, which is the usual tight-binding chain, already studied under other monitoring protocols \cite{chahineEntanglementPhasesLocalization2024,poboikoNonlinearSigmaModels2023}. We expect to retrieve the same symmetry class. This limit means that we consider $\eta = 0$ and $J\neq 0$, $\gamma\neq 0$.

We find $N_f = 2R^2 $ free parameters in Table~\ref{tab:no_pairing}, a), where antisymmetric matrices have $(R^2 - R)/2 $ free parameters, whereas symmetric one have $(R^2 + R)/2$ ones. This number corresponds to the group $G = U(R) \times U(R)$. Additionally, to obtain the NLSM manifold, the rotation should be a non trivial rotation of the saddle point $\Lambda$ to represent a meaningful fluctuation. Hence, we should consider the group $H$ leaving the saddle point invariant, and our final manifold will be the quotient $G/H$. 

Since the saddle point reads as $Q_s = \mathbbm{1}_R \otimes \sigma_z \otimes \tau_0 $, we obtain the structure of Table~\ref{tab:no_pairing}, b), which gives $N_f = R^2$ free parameters and a group $H = U(R)$. Finally, we find the NLSM manifold $ G/H = U(R)$ from these constraints, and since $\mathrm{det} \left[ Q \right] = 1$ we obtain
\begin{equation}
    G/H = SU(R),
\end{equation}
the special unitary group, corresponding to the AIII class.
This is indeed the result of the analysis of Ref.~\cite{poboikoNonlinearSigmaModels2023} for the case of free fermions with density monitoring. We note that in this case there is a non-trivial symmetry in the problem also in the replica limit $R\rightarrow 1$, corresponding to $U(1)$ rotations associated to the conservation of particle number at the level of the average state. 

\subsubsection{General Ising Chain}
\label{sec:class_diii}

\begin{table}[h]
    \centering
    \begin{tikzpicture}
        \node (array1) {
            $\begin{tabular}{||c|| c c c c||}
                \hline
                & $\sigma_0$ & $\sigma_x$ & $\sigma_y$ & $\sigma_z$ \\ [0.5ex]
                \hline\hline
                $\tau_0$ & A,R & A,R  &   & \\
                \hline
                $\tau_x$ &  &  &   &   \\
                \hline
                $\tau_y$ &   &  &   &   \\
                \hline
                $\tau_z$ &  &   &  &  \\
                \hline
            \end{tabular}  $
        };
        \node[left=0.3cm of array1, yshift=1cm] {a)};

        \node (array2) at (7.5cm,0) {
            $\begin{tabular}{||c|| c c c c||}
                \hline
                & $\sigma_0$ & $\sigma_x$ & $\sigma_y$ & $\sigma_z$ \\ [0.5ex]
                \hline\hline
                $\tau_0$ & A,R &   &   &  \\
                \hline
                $\tau_x$ &  &  &   &  \\
                \hline
                $\tau_y$ &   &  &   &   \\
                \hline
                $\tau_z$ &   &   &  &  \\
                \hline
            \end{tabular} $
        };
        \node[left=0.3cm of array2, yshift=1cm] {b)};
    \end{tikzpicture}
    \caption{ Case of the general Ising chain : $\eta \neq 0$, $J\neq 0$,  and $\gamma\neq 0$ - a) Structure of the $W_{ij}$ matrice under constraints $(\boldsymbol{C0})$, $(\boldsymbol{C1})$, $(\boldsymbol{C2})$, $(\boldsymbol{C3})$ and  $(\boldsymbol{C4})$, b) Structure of the $W_{ij}$ under the constraints of a) and saddle point invariance additionally. }
    \label{tab:full}
\end{table}

We now consider the general case with pairing terms, relevant for the monitored Ising chain of our interest, and follow the same strategy ($\eta \neq 0$, $J\neq 0$, and $\gamma\neq 0$). We find for this enlarged symmetry $N_f = 2 R(R-1)/2$ free parameters, which corresponds to the group $G = O(R) \times O(R)$. This is ultimately due to the fact that the pairing term $\eta$ introduces an additional constraint which reduces the dimension of the symmetry group. If we add the saddle point invariance in the constraints, one obtains $N_f =  R(R-1)/2$ leading to the group $H = O(R)$. Finally we get the NLSM manifold $G/H = O(R)$ which again with the constraint on the determinant gives 
\begin{equation}
    G/H = SO(R),
\end{equation}
the special orthogonal group, which corresponds to the DIII class. This is consistent with previous findings in a Majorana chain with random couplings within the context of quantum state diffusion in Ref.~\cite{favaMonitoredFermionsConserved2024}. We note that in this case, as opposed to the previous one, the $R \rightarrow 1$ limit is trivial from the point of view of symmetries. Indeed, the pairing term $\eta$ breaks explicitly the strong symmetry associated to particle number conservation, and prevents the average state to display diffusive dynamics of the charge sector.

\subsubsection{Pairing Limit of the Ising Chain}

To complete the exploration of the symmetries across the phase diagram of the monitored Ising chain we now finally consider the special limit where the tight-binding term is absent, and only the pairing is present in the Hamiltonian ($J= 0$ and $\eta \neq 0$,  and $\gamma \neq 0$). This corresponds to a monitored Ising chain with $J_x+J_y=0$ in Eq.~(\ref{eq:IsingChain}).

\begin{table}[b]
    \centering
    \begin{tikzpicture}
        \node (array1) {
            $\begin{tabular}{||c|| c c c c||}
                \hline
                & $\sigma_0$ & $\sigma_x$ & $\sigma_y$ & $\sigma_z$ \\ [0.5ex]
                \hline\hline
                $\tau_0$ & A,R & A,R  &   & \\
                \hline
                $\tau_x$ &  &  &   &  \\
                \hline
                $\tau_y$ &   &  & A,R  & S,I   \\
                \hline
                $\tau_z$ &  &   &  &  \\
                \hline
            \end{tabular}  $
        };
        \node[left=0.3cm of array1, yshift=1cm] {a)};

        \node (array2) at (7.5cm,0) {
            $\begin{tabular}{||c|| c c c c||}
                \hline
                & $\sigma_0$ & $\sigma_x$ & $\sigma_y$ & $\sigma_z$ \\ [0.5ex]
                \hline\hline
                $\tau_0$ & A,R &   &   &  \\
                \hline
                $\tau_x$ &  &  &   & \\
                \hline
                $\tau_y$ &   &  &   & S,I   \\
                \hline
                $\tau_z$ &   &   &  &  \\
                \hline
            \end{tabular} $
        };
        \node[left=0.3cm of array2, yshift=1cm] {b)};
    \end{tikzpicture}
    \caption{Case of the pairing limit of the Ising chain : $J= 0$ and $\eta \neq 0$,  and $\gamma\neq 0$ - a) Structure of the $W_{ij}$ matrice under constraints $(\boldsymbol{C0})$, $(\boldsymbol{C1})$, $(\boldsymbol{C2})$ and  $(\boldsymbol{C4})$, b) Structure of the $W_{ij}$ under the constraints of a) and saddle point invariance additionally. }
    \label{tab:dual_J}
\end{table}

For that case the number of free parameters is $N_f = 2R^2 -2 $, which corresponds to the group $G = O(2R)$. This is due to the fact that in absence of hopping term in the Hamiltonian the structure in Keldysh-Nambu changes and so there is one less constraint to impose from Eq.~(\ref{eq:constraints_C2_C4}). If we add the saddle point invariance in the constraints, we obtain the Table~\ref{tab:dual_J}, b) and $N_f =  R^2$ leading to the group $H = U(R) $. This yields the NLSM manifold $G/H = O(2R)/U(R)$ which restricts to (with the determinant constraint)
\begin{equation}
    G/H = SO(2R)/U(R),
\end{equation}
which corresponds to the D class~\cite{eversAndersonTransitions2008}. We note that a similar symmetry class was found in the case of noisy Majorana fermions with diffusive monitoring along the measurement-only line~\cite{favaNonlinearSigmaModels2023}. Here instead the D symmetry class appears in the pairing limit of the monitored Ising chain, where next-neighbor pairing competes with local density monitoring. Numerical investigations of this regime have not been performed to the best of our knowledge. We note however that, as anticipated, in this limit the monitored Ising chain can be mapped into a monitored Su-Schrieffer-Heeger model~\cite{legalEntanglementDynamicsMonitored2024} under particle-hole transformation of fermions on the even sites. This raises the intriguing question, which we leave for future studies, of what is the field theory description for this model.

\subsubsection{Method 2 : Finding the Rotation Matrix $\mathcal{R}$ }
\label{sec:rotation_mat_param}

In this section, we directly find a parametrisation of the matrix $\mathcal{R}$ by imposing the constraints. While this method is less automatic than the previous approach—particularly when the expected outcome is unclear— but it proves valuable in deriving the NLSM due to its explicit formulation, as will be demonstrated in the following section. We here only consider the general case $\eta \neq 0$, $J\neq 0$, and $\gamma\neq 0$.\\

We start by writing the rotation matrix as 
\begin{equation}
    \mathcal{R}  = \begin{pmatrix}
        A & B \\
        C & D 
    \end{pmatrix}, \quad \mathrm{with} \quad A,B,C,D \in \mathcal{M}(2R).  
\end{equation}
The constraint $(\boldsymbol{C2})$, due to the measurement terms impose that $B  = \sigma_z C \sigma_z$ and $ A = \sigma_z D \sigma_z $, which allows us to write 
\begin{equation}
    \mathcal{R} = \begin{pmatrix}
        A & B \\
        \sigma_z B \sigma_z  & \sigma_z A \sigma_z 
    \end{pmatrix}.
\end{equation}
The condition $(\boldsymbol{C3})$ reading as $ \left[ \mathcal{R} , \mathbbm{1}_R \otimes \sigma_0 \otimes \tau_z \right]  = 0$, imposes $A= \sigma_z A \sigma_z$ and $B = \sigma_z B \sigma_z$ which means that the matrices $A$ and $B$, now reads as 
\begin{equation}
   A = \begin{pmatrix}
        a_1 & 0 \\
        0 & a_2
    \end{pmatrix} \quad \mathrm{and} \quad B = \begin{pmatrix}
        b_1 & 0 \\
        0 & b_2
    \end{pmatrix},
\end{equation}
where $a_1$, $a_2$, $b_1$ and $b_2$ are $R\times R$ matrices. The constraint $(\boldsymbol{C1})$,  $\mathcal{C} \mathcal{R}^* \mathcal{C} = \mathcal{R}$ imposes that $\sigma_y A^* \sigma_y =A$  and $\sigma_x B^* \sigma_x =B $, and $(\boldsymbol{C4})$ gives  $\sigma_y A \sigma_y =A$  and $\sigma_x B \sigma_x =B$ which implies that $A$ and $B$ are real. Moreover, from $(\boldsymbol{C4})$, we infer $\sigma_x A \sigma_x =A$ and $\sigma_x B \sigma_x =B$, implying that $a \equiv a_1 = a_2$ and $b \equiv b_1 = b_2$. Finally, the unitary condition entails that
\begin{equation}
    a^T a + b^T b =  \mathbbm{1}_R \quad \mathrm{and} \quad b^T a + a^T b = 0,
\end{equation}
from which we deduce that $(a+b)$ and $(a-b)$ both belong to $O(R)$. Therefore adding the determinant constraint, the rotation matrix $\mathcal{R}$ belongs to $SO(R)\times SO(R)$ except when $a+b = a-b$ in which case $\mathcal{R}$ belongs to $SO(R)$. We will see later that this last case happens when the rotation matrix is leaving the saddle point invariant, we thus retrieve the same result as in Section \ref{sec:class_diii}, with the DIII class.

\subsection{Derivation of the Non-Linear Sigma Model}\label{sec:non_liner_sigma}

In this section, we explicitly derive a Non Linear Sigma Model using an approach similar to that of Ref.~\cite{poboikoMeasurementinducedTransitionsInteracting2025}, adapted to account for the specific features of the Ising chain. 

\subsubsection{Rotation Matrix Parametrisation}
\label{ch6:sec:rotation_param}

We have shown in the previous section that the rotation matrix $\mathcal{R}$ can be parametrised as 
\begin{equation}
    \mathcal{R} = \begin{pmatrix}
       A & B \\
       B & A
    \end{pmatrix}, \quad \mathrm{with} \quad A =\begin{pmatrix}
        a & 0 \\
        0 & a 
    \end{pmatrix} , \quad B =\begin{pmatrix}
        b & 0 \\
        0 & b 
    \end{pmatrix}, 
\end{equation}
where $a$ and $b$ are $R\times R$ matrices, such that $a+b$ and $a-b$ belongs to $SO(R)$. By performing a change of basis, we bring this matrix in a diagonal form such that 
\begin{equation}
    \mathcal{R}^\prime = \mathcal{P}^T \mathcal{R} \mathcal{P} = \begin{pmatrix}
        \mathcal{V}_+  & 0 \\
        0 & \mathcal{V}_-
    \end{pmatrix}_K \otimes \tau_0  ,\quad  \mathcal{V}_+, \mathcal{V}_- \in SO(R)
    \label{eq:rotation_param}
\end{equation}
with $\mathcal{V}_+ = a+b$ and $\mathcal{V}_-=a-b$. This rotation corresponds to operate in the basis $\mathcal{B}^\prime_{KN}$, which reads as\footnote{ Indeed since $Q=\mathcal{R} Q \mathcal{R}^{-1}$ we find that the new basis should satisfy $\Psi^\prime = \mathcal{P}^T \Phi$} 
\begin{align}
    \Psi_{\mathrm{KN},r}^{\prime T} &= \begin{pmatrix}
        \psi_{+,x,r} & \overline{\psi}_{+,x,r} & -\psi_{-,x,r} & \overline{\psi}_{-,x,r}
    \end{pmatrix} / \sqrt{2}\\   \overline{\Psi}^\prime_{\mathrm{KN},r} &=\begin{pmatrix}
        \overline{\psi}_{+,x,r}    & \psi_{+,x,r} & \overline{\psi}_{-,x,r}& -\psi_{-,x,r}
    \end{pmatrix} / \sqrt{2},
    \label{ch6:eq:basis_rotation}
\end{align}
meaning that 
\begin{equation}
    \overline{\Psi}_{\mathrm{KN},r}^{\prime T} =  (\sigma_0 \otimes \tau_x ) \Psi_{\mathrm{KN},r} \quad \mathrm{and} \quad  \rho = \frac{1}{2} - \frac{1}{8} \mathrm{Tr} \left[ Q_0 (\sigma_z \otimes \tau_z) \right].
\end{equation}  



\subsubsection{NLSM Derivation}
\label{ch6:sec:non_liner_sigma}

We recall here for convenience the form of the bosonic action Eq.~\eqref{eq:full_action} reading as \footnote{Where we have dropped the constant factor 2 in the logarithm since constants are not playing role since they must globally sum to zero to have a Keldysh partition function equal to unity in the limit $R \rightarrow 1$  } 
\begin{equation}
   \nonumber S = -i \mathrm{Tr}\left[\frac{1}{2}\ln\left( -i G_{0,\mathrm{KN}}^{-1} + \Sigma \right) \right] -\mathrm{Tr}\left[\Sigma \mathcal{G}\right] -  i \gamma \int dt dx ~ \frac{1}{4} \sqrt{\mathrm{det}[\mathbb{1}_4 - 4i \mathcal{G}(x,t) (\mathbb{1}_R \otimes \sigma_z \otimes \tau_z) ]},
\end{equation}
where we now adopt the basis $\mathcal{B}^\prime_\mathrm{KN}$ that simplifies the derivation, we from now on call $\mathcal{R}$ the matrix $\mathcal{R}^\prime$ of Eq.~\eqref{eq:rotation_param}. This changes the matrices representation, for example $\sigma_x \otimes \tau_z  \rightarrow \sigma_z \otimes \tau_z  $. The goal of the following computation is to derive a Non Linear Sigma Model (NLSM), indeed, we found a saddle point and we now seek for the full manifold of matrices that do not incur any additional energy cost. These modes are called the massless or Goldstone modes, and are at low energy the relevant one since they have long range correlations (contrary to the massive one, which usually are decaying exponentially due to the mass and can be integrated out). The term $\mathrm{Tr}\left[\Sigma \mathcal{G} \right]$ of the action is proportional to $\mathrm{Tr}\left[Q^2 \right]$, and since we found that $\Lambda^2 = \mathbb{1}_4$, to preserve the energy, a reasonable choice is to impose that $Q^2 = \mathbb{1}_{4R}$. This non-linear constraint imposes that any $Q$ matrix in this manifold can be parametrised as 
\begin{equation}
    Q = \mathcal{R}^{-1} \Lambda \mathcal{R},
\end{equation} 
where $ \mathcal{R}$ is a unitary matrix\footnote{Since $Q$ is an involution with $\mathrm{Tr}\left[ Q \right] = 0$, $Q$ and $\Lambda$ have the same spectrum with $2R$ eigenvalues $1$ and $2R$ eigenvalues $-1$. We can therefore find $Q$ by performing a rotation of $\Lambda$ with $\mathcal{R}$. }. In order to describe massless fluctuations, the transformation $\mathcal{R}$ must respect the symmetries previously mentionned which leave the action invariant, therefore we can employ parametrisation of $\mathcal{R}$ previously established in Section \ref{ch6:sec:rotation_param}. Remarkably, and as noted previously, the rotation matrix will only act in the replica space, since no symmetry is left at the level of the $\mathrm{Keldysh} \otimes \mathrm{Nambu}$ space, as we have discussed in Section \ref{sec:class_diii}.

To derive the NLSM, we will only consider slowly varying in space and time rotation matrices $\mathcal{R}(x,t)$, and perform a gradient expansion to obtain the final result. 
In the action Eq.~\eqref{eq:action_full_KN_real}, the measurement term is invariant on the NLSM manifold, and only gives a constant that is irrelevant since constants should sum to zero in the limit $R \rightarrow 1$ \cite{poboikoMeasurementinducedTransitionsInteracting2025}. The term $\mathrm{Tr}\left[\Sigma \mathcal{G} \right]$ because of the non-linear constraint is also giving a constant and thus ignored. We finally focus on the fermionic determinant, which reads by using $\Sigma = \tfrac{\gamma}{2} \left( Q - \mathbbm{1}_R \otimes  \sigma_z \otimes \tau_z \right)$ and $\mathcal{G} = - i Q /4 $ (found in Section \ref{sec:ising_saddle}) as\footnote{In this expression we are already taking the limit $R \rightarrow 1$ in the prefactors for clarity, the complete expression can be found in Appendix \ref{app:nlsm_deriv} }
\begin{equation}
    S[Q] = -\frac{i}{2}\mathrm{Tr}\left[ \ln\left( -i g_{0,\mathrm{KN}}^{-1} + \frac{\gamma}{2} \mathcal{R} \Lambda \mathcal{R}^{-1} \right) \right] ,
\end{equation}
where $S[Q]$ can be understood as $S[\mathcal{R}, \Lambda]$. For convenience in the following we drop the $\mathrm{KN}$ subscript. Since $g_{0}^{-1} = i \partial_t - H$, we define 
\begin{align}
    W &\equiv \mathcal{R}^{-1} [-ig_{0}^{-1}, \mathcal{R}]\equiv  \mathcal{R}^{-1} [\partial_t , \mathcal{R}]  + i \mathcal{R}^{-1} [H_0 , \mathcal{R}]  \approx \mathcal{R}^{-1} \partial_t \mathcal{R}+ \hat{v} \mathcal{R}^{-1} \nabla_x \mathcal{R},
\end{align}
and we note $W^{(x)} =  \hat{v} \mathcal{R}^{-1} \nabla_x \mathcal{R}$ and $W^{(t)} =   \mathcal{R}^{-1} \partial_t \mathcal{R}$. We thus obtain by factorizing by the dressed Green's function  $G^{-1} = g_0^{-1} + i \tfrac{\gamma}{2} \Lambda$, 
\begin{align}
    S[Q] &=  -\frac{i}{2} \mathrm{Tr}\left[\ln\left( -i G^{-1}    \right) \right]  -\frac{i}{2} \mathrm{Tr}\left[\ln\left( 1 + i G W   \right) \right].
\end{align}
We note that $H_0$ is the Hermitian Hamiltonian since the non-Hermitian part cancels with the expression of the self energy $\Sigma$. By performing a Taylor expansion of the logarithm at second order in $W$ one gets 
\begin{equation}
    S[Q] \approx \frac{1}{2} \mathrm{Tr}\left[ G W  \right]  - \frac{i}{4} \mathrm{Tr}\left[ G W GW  \right].
\end{equation} 
At first order the spatial term vanishes since it is linear in $v(k) = \partial_k \xi_k$, which is antisymmetric and the temporal term can only sustain a replica symmetric term\footnote{This essentially comes from  $\mathrm{Tr}_R \left[ \mathcal{R}^{-1} \partial_t  \mathcal{R} \right]$ which is a matrix composed of terms $\mathrm{Tr}_R[\mathcal{V}_\pm ^{-1}\partial_t \mathcal{V}_\pm]$. Since $\mathrm{det}[\mathcal{V}_\pm] = 1 $, the derivative is null and since $ \partial_t \mathrm{det}[\mathcal{V_\pm}] = \mathrm{det}[\mathcal{V}_\pm] \mathrm{Tr}_R[\mathcal{V}_\pm ^{-1}\partial_t \mathcal{V}_\pm]$,  we find $\mathrm{Tr}_R[\mathcal{V}_\pm ^{-1}\partial_t \mathcal{V}_\pm]= 0$.}, which does not exist in the this case with the $SO(R)$ symmetry. Hence, we do not find any linear contribution. 

  
We now consider the second order terms, and by considering the parametrisation $ G = \frac{1}{2} G^R (1+ \Lambda) + \frac{1}{2} G^A (1 - \Lambda)$, we can show that only cross terms proportional to $G^R G^A $ are contributing, resulting for the spatial contribution in  
\begin{align}
    & S[Q]^{(2,2)}= -\frac{i}{4} \tr{GW^{(x)}GW^{(x)} } = \frac{i}{8} D ~\mathrm{Tr}\left\{ \left[\nabla_x Q \right]^2 \right\},
\end{align}
where the diffusion coefficient $D$, is defined through the group velocity $v(k) = \partial_k \xi_k$ as 
\begin{equation}
   D =   \int_{-\pi}^{\pi} \frac{dk}{2\pi}~ \frac{v(k)^2}{\gamma} = \frac{v_0^2 }{\gamma} .
\end{equation}
By doing a similar computation for the temporal part of the action, we obtain a NLSM under the form 
\begin{equation}
    S[Q] = \frac{i}{8} \frac{v_0^2}{\gamma} ~\mathrm{Tr}\left\{ \left[\nabla_x Q \right]^2  + \frac{1}{v_0^2}\left[\partial_t Q \right]^2 \right\}. 
\end{equation}
Our goal is to single out the replica contribution, this is done by exploiting the structure of $\mathcal{R}$ as shown in Eq.~\eqref{eq:rotation_param} and the replica symmetric structure of $Q_0=\Lambda$. By writing that $Q = \mathcal{R} Q_0 \mathcal{R}^{-1}$, and $U= \mathcal{V}_+ \mathcal{V}_-^T$, we obtain after lengthy algebra detailed in Appendix \ref{app:nlsm_deriv} that 
\begin{equation}
    S[U] =   \frac{2i\rho(1-\rho) v_0}{\gamma} ~  \mathrm{Tr}_R\left\{  \frac{1}{v_0} \partial_t U^T \partial_t U + v_0 \nabla_x U^T \nabla_x U  \right\} 
\end{equation}
with  $\rho = \frac{1}{2} - \frac{1}{8} \mathrm{Tr} \left[ Q_0 (\sigma_z \otimes \tau_z) \right]$.


\section{Discussion}\label{sec:discussion}

In the previous sections we have obtained the replica field theory for the monitored Ising chain under the QJs protocol. By using the parameterization found in Section \ref{sec:rotation_mat_param} we have derived an effective field theory describing the slow modes of this replicated system. This effective model takes the form of a Non Linear Sigma Model usually found in the context of disordered systems, and reads as 
\begin{equation}
    S[U] =  \frac{1}{g_B} \mathrm{Tr}\left[ \partial_\mu U^T \partial_\mu U \right],
\end{equation}
where $\mu=v_0t, x$ is summed over and corresponds to time and space indices, and $U\in SO(R)$ and $g_B$ the bare coupling constant. We stress that these matrices are of size $R \times R$ and thus describes the replicon sector (ie the replica symmetric transverse space). As we showed explicitly in Section \ref{sec:symmetries}, this symmetry class corresponds in the ten fold classification \cite{eversAndersonTransitions2008,ludwigTopologicalPhasesClassification2016} to the DIII class, which belongs to the Bogoliubov-De-Gennes category. 
The knowledge of the symmetry class is crucial to assess the emergent low-energy, large-scale physics of the NLSM. Indeed, the perturbative renormalisation group (RG) flow of the NLSM, describing this regime, depends of the different symmetry classes. These flows have been computed for any integer value of $R$ in Ref. \cite{hikamiThreeloopBetafunctionsNonlinear1981} (see also Ref.~\cite{eversAndersonTransitions2008}) and in our case of the DIII class case, we find
\begin{equation}
    \frac{d g_R}{d \ln L } = \frac{1}{8\pi} (R-2) g_R^2 + O(g^3_R ),
    \label{eq:betafunctionDIII}
\end{equation}
where $g_R$ is the renormalised coupling constant. 

We need to remember that to conclude about the physics we are interested in we should take the replica limit $R\rightarrow 1$. In that case, we observe that the flow of the coupling constant is towards weak coupling, i.e. $g_R=0$ is a fixed point. This implies that perturbation theory in the monitoring strength is well behaved in our problem, and that the behavior of the $\gamma/J\ll 1$ regime, where the NLSM derivation is controlled, represents a stable phase of the monitoring Ising chain.

Our result is compatible with the analysis of Ref.~\cite{favaNonlinearSigmaModels2023}, which considered a very different microscopic model, involving noisy Majorana fermions undergoing continuous time (quantum state diffusion) monitoring, yet sharing the same fundamental symmetries. The derivation of the NLSM between the two approaches was rather different: in our case, the lack of noise in the unitary evolution required us to use Keldysh techniques as done in the theory of Anderson localization.
We emphasize that the NLSM result does not tell much about the regime of strong monitoring, which is outside the range of validity of our treatment ($\gamma \ll 1$). It only ensures the weak-monitoring phase is stable, but cannot provide details about the transition out of this phase. Based on perturbative arguments and numerical evidence\cite{turkeshiMeasurementinducedEntanglementTransitions2021, legalEntanglementDynamicsMonitored2024}, it is natural to expect that the strong monitoring regime is dominated by a stable Zeno phase characterized by an area-law scaling of entanglement entropy, which corresponds to a disordered phase of the NLSM.

Our approach has the advantage to naturally connect with the literature on monitored free fermions with strong $U(1)$ symmetry. Indeed as discussed in Sec.~\ref{sec:symmetries}, our monitored Ising chain reduces to free fermions in absence of pairing in the Hamiltonian, i.e. $\eta=0$, and we have indeed shown that the symmetry class of the effective NLSM in that case reduces to 
$SU(R)$ symmetry corresponding to the AIII class \cite{poboikoNonlinearSigmaModels2023,chahineEntanglementPhasesLocalization2024,favaMonitoredFermionsConserved2024}. We also expect, even though we have not shown it explicitly, that $\eta=0$ and in presence of particle-hole symmetry our framework would correctly give the $SU(2R)/Sp(2R)$ symmetry characteristic of the BDI class \cite{poboikoMeasurementinducedTransitionsInteracting2025,favaMonitoredFermionsConserved2024}. In both cases the low-energy physics of the resulting NLSM is different from Eq.~\eqref{eq:betafunctionDIII}: in the replica limit $R \rightarrow 1$ the flow is towards strong coupling, indicating that the weak monitoring phase is unstable at large dissipation, via a mechanism similar to the weak-localization corrections to Anderson localization~\cite{poboikoNonlinearSigmaModels2023}. This crucial difference in symmetry class has, therefore, the important consequence that free fermions with strong $U(1)$ symmetry flow always towards an area law phase of the entanglement entropy, even if the length scale controlling this crossover can be very large and undetectable in finite-size numerical simulations. Our results, on the other hand, show that the monitored Ising chain, in fact for any small value of the pairing term $\eta\neq0$, displays a robust entangled phase and thus a genuine measurement-induced phase transition into a Zeno area-law phase.

\section{Conclusions}\label{sec:conclusions}

In this work we have derived the replica description of monitored quantum systems evolving under the quantum jumps protocol. 

Our first result is that the replicated master equation for the QJs protocol involve a non-Hermitian evolution, which is diagonal in replica space and a coupling term between replicas arising only from the quantum jumps. This is true for generic Hamiltonian and monitoring operator, irrespective of whether the system is interacting or not.  

We have then converted this master equation in a Keldysh field theory and specialised our analysis to the monitored Ising chain, for which we have written the replica Keldysh field theory and obtained the bosonic effective action in terms of slowly varying fields. From this we have derived the saddle point equations and discussed how they reproduce the physics of the averaged density matrix evolving under Lindblad dynamics.

We have then discussed in details the overall symmetries of the replica field theory. Based on this analysis we have derived the effective NLSM field theory describing the slow degrees of freedom, the Goldstone modes associated to the replica symmetry. We have shown that for the Ising chain the NLSM corresponds to the DIII symmetry class, except at a special point in the phase diagram - corresponding to $J_x+J_y=0$ where hopping is absent and only next-neighbor pairing is present in the fermionized Hamiltonian - which corresponds to the D class. Finally, using the results obtained we have discussed the stability of the weak-monitoring phase for the Ising chain and of the MIPT. 

Our work opens up a number of directions to explore in the near future. One could consider different measurement operators such as particle losses, studied numerically in Ref.~\cite{soaresEntanglementTransitionDue2025} which reported an entanglement transition similar to one found for the density monitoring. At the level of field theory losses would result in a different structure of the quantum jump term in the Keldysh action, as discussed in the case of free fermions with $U(1)$ (weak) symmetry in Ref.~\cite{starchlGeneralizedZenoEffect2024}, whose impact on the field theory is worth investigating. One could also consider measurement operators which break the Gaussianity of the non-Hermitian Hamiltonian and that would need to be treated on the same footing as quantum jumps.

Similarly, one could consider different Hamiltonian - such as the limit $J=0$ of the Ising chain where the symmetry class changes - or include interactions in the Hamiltonian - such as second neighbor Ising coupling - and discuss their role, in a case where the replica symmetric sector is not coupled to a diffusion mode, as opposed to Refs.~\cite{poboikoMeasurementinducedTransitionsInteracting2025,guoFieldTheoryMonitored2024}. Another direction worth exploring with our approach would be to include quenched disorder, which is known to have strong impact on both non-Hermitian physics and monitoring.

Finally, one could try to explore further the role of the non-Hermitian Hamiltonian and the associated symmetry classes~\cite{kawabata2019symmetry,lieu2020tenfold,sa2023symmetry} in determining the universality of MIPT. Indeed, so far only the symmetry classes corresponding to the (unitary) ten-fold way classification have been identified~\cite{bhuiyan2025freefermiondynamicsmeasurementstopological}, yet monitored systems are fundamentally non-unitary field theories which might open up to new scenarios~\cite{xiao2025topologymonitoredquantumdynamics}. We leave these exciting questions for future works.

\emph{Note Added}: Upon completion of our work we became aware of Ref.~\cite{foster2025freefermionmeasurementinducedvolumearealaw} which also consider the monitored Ising chain, although with a different monitoring protocol. Our results are compatible with their analysis, when they overlap.

\section*{Acknowledgements}
Y.L.G. acknowledges useful discussions with M. Vanhoecke.
\paragraph{Funding information}
This work has received funding from the European Research Council (ERC) under the European Union's Horizon 2020 research and innovation programme (Grant agreement No. 101002955 -- CONQUER)
The  computations  were  performed  on  the  Collège de France IPH computer cluster.

\begin{appendix}

\section{Time Integrated Replicated Master Equation}

\label{app:time_int_master_eq}

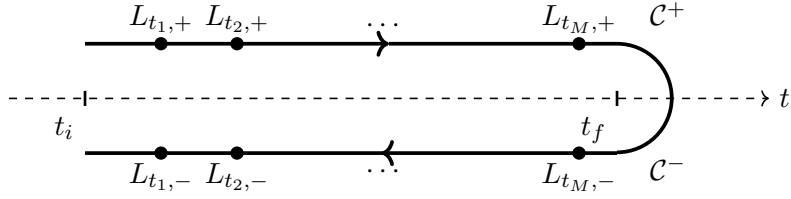
\begin{figure}[h]
    \centering
    \begin{tikzpicture}
        \draw[line width=1.4pt][->] (-1,0.1) -- (3,0.1) node[right] {};
        \draw[line width=1.4pt] (3,0.1) -- (6,0.1) node[right] {};
        
        \draw[line width=1.4pt][->] (6,-1.35) -- (2.9,-1.35) node[right] {};
        \draw[line width=1.4pt] (3,-1.35) -- (-1,-1.35) node[right] {};

        \draw (6,0.1)[line width=1.4pt]  arc (90:-90:0.72cm);

        \draw[line width=0.6pt, dashed][->] (-2,-0.625) -- (8,-0.625) node[right] {$t$}; 

        \fill (0,0.1) circle (2.5pt) node[above] {$L_{t_1,+}$}; 
        \fill (1,0.1) circle (2.5pt) node[above] {$L_{t_2,+}$}; 
        \fill (5.5,0.1) circle (2.5pt) node[above] {$L_{t_M,+}$}; 
        \node at (2.95,0.12) [above] {$\cdots$};

        \fill (0,-1.35) circle (2.5pt) node[below] {$L_{t_1,-}$}; 
        \fill (1,-1.35) circle (2.5pt) node[below] {$L_{t_2,-}$}; 
        \fill (5.5,-1.35) circle (2.5pt) node[below] {$L_{t_M,-}$}; 
        \node at (2.95,-1.37) [below] {$\cdots$};

        \draw[line width=1pt] (-1,-0.625 + 0.1) -- (-1,-0.625 -0.1) node[below left] {$t_i$};

        \draw[line width=1pt] (6,-0.625 + 0.1) -- (6,-0.625 -0.1) node[below left] {$t_f$}; 
        \node at (6.3,0.5) [right] {$\mathcal{C}^+$};
        \node at (6.3,-1.55) [right] {$\mathcal{C}^-$}; 
    \end{tikzpicture}

    \caption{Scheme of the considered contour $\mathcal{C}$}
    \label{fig:contour}
\end{figure}

It is possible to derive the master equation \eqref{eq:replicated_master_eq} in a time integrated way, instead of working with the time-step evolution $t+dt$ as done before. We start from Eq.~\eqref{eq:average_traj_replica}, where we now perform the average conditionally to the number of jumps $M$ happening during the trajectory. We obtain (omitting as before the partial trace and the limit $R \rightarrow1$)
\begin{equation}
    \rho^R_t = \sum_{M=0}^\infty \int_{\{x_m, t_m\}} \left( \bigotimes_{r=1}^R \left[ S(t_f,t_i) \rho_0   S(t_f,t_i)^\dagger \right] \right), 
\end{equation}
where $\int_{\{x_m, t_m\}}=  \int_{t_i < t_1< ... < t_{M} < t_f } d\mathbf{t}  \sum_{x_1,..., x_M=0}^L$  and $S(t_f,t_i) =  U_\mathrm{nH}(t_f,t_M) L_{x_M}  ...   L_{x_{1}} U_\mathrm{nH}(t_1,t_i)$. We recall that $ \rho^R_t$ stands for $ \overline{\rho}_t^{\otimes R}$ with $\overline{\square}$ the unweighted sum over trajectories. Using the contour depicted in Fig.~\ref{fig:contour} and defined as $\mathcal{C} = \mathcal{C}^+ \cup \mathcal{C}^-= (t_i, t_f) \cup (t_f, t_i)$, we obtain under contour ordering $\mathbb{T}_\mathcal{C}$,
\begin{equation}
    \rho^R_{t_f} = \mathbb{T}_\mathcal{C} \left[ \left( U_\mathrm{nH}(t_f,t_i)  \rho_0  U^\dagger_\mathrm{nH}(t_f,t_i) \right)^{\otimes R} \sum_{M=0}^\infty   \frac{1}{M! } \left( \gamma \int_{t_0}^{t_f} dt \sum_{x = 1}^L \left[ L_{x , (t,+)} L^{\dagger}_{x , (t,-)}  \right]^{\otimes R } \right)^M  \right].
\end{equation} 
The factor $1/M!$ is added to avoid multiple counting since we are not ordering the times $t_i$ in the integral anymore. We also attach a fictional time and contour symbol $+\-$ to the jump operator to be able to perform the contour ordering in the end. By noticing the expansion of the exponential and using $H_\mathrm{eff} = H - i \tfrac{\gamma}{2} \sum_{x=1}^L L_x^\dagger L_x$ we find 
\begin{equation}\label{eq:replicated_rho}
    \rho^R_{t_f} = \mathbb{T} \left[ \exp\left(-i \int_\mathcal{C} dt \sum_{r=1}^R H_\mathrm{eff,r}(t) \right)  \exp\left( \gamma \int_{t_i}^{t_f} dt  \sum_{x=1}^L  \prod_{r=1}^R L_{x , (t,+)}^{(r)} L^{(r)\dagger}_{x ,  (t,-)}  \right)  \rho_0^{\otimes R } \right]
\end{equation}
which again describe the same evolution as discussed previously.

\section{ Regularisation of Green's Functions  }
\label{app:regularisation_green_function}

In this Appendix, we discuss the regularisation of the Green's function when we are considering local in time interactions. 

\paragraph{The Green's function discontinuity:}First in Keldysh formalism we have two main basis, the $+/-$ basis where 
\begin{equation}
    -\left\langle \psi_m (t) \overline{\psi}_n (t^\prime) \right\rangle = G_{mn}(t,t^\prime) = \begin{pmatrix}
        G^{\mathbb{T}}(t,t^\prime) &  G^{<}(t,t^\prime) \\
         G^{>}(t,t^\prime) & G^{\tilde{\mathbb{T}}}(t,t^\prime)
    \end{pmatrix},
\end{equation}
with $m,n=(+,-) $ and $G^{<}(t,t^\prime)$ the lesser component, $G^{>}(t,t^\prime)$ the greater component and $ G^{\mathbb{T}}(t,t^\prime)$ (resp. $G^{\tilde{\mathbb{T}}}(t,t^\prime)$) the time orderding (resp. time-ordering components). In the Larkvin-Ovchinikov (LO) basis we have 
\begin{equation}
    - i \left\langle \psi_a (t) \overline{\psi}_b (t^\prime) \right\rangle = G_{ab}(t,t^\prime)= \begin{pmatrix}
        G^R(t,t^\prime) & G^K(t,t^\prime)  \\
        G^{\overline{K}}(t,t^\prime) & G^A(t,t^\prime)  \\
    \end{pmatrix},
\end{equation}
where $a,b = (1,2)$, and the $R$ denotes the retarded component, $A$ the advanced component and $K$ (resp. $\overline{K}$) the Keldysh component (resp. the anti-Keldysh component). Using the LO definition 
\begin{equation}
    \psi_{1,2} = \frac{1}{2}\left(\psi_+(t) \pm \psi_- (t) \right) \quad \mathrm{and} \quad \overline{\psi}_{1,2} = \frac{1}{2}\left(\overline{\psi}_+(t) \mp \overline{\psi}_- (t) \right),
\end{equation}
we can express the Keldysh and anti-Keldysh components
\begin{equation}
    \begin{cases}
        G_{\bar{K}}(t, t^\prime) &=   \frac{1}{2} \left[ G^\mathbb{T}(t, t^\prime) - G^>(t, t^\prime) - G^<(t, t^\prime) + G^{\tilde{\mathbb{T}}}(t, t^\prime) \right]  \\
         G_{K}(t, t^\prime) &=   \frac{1}{2} \left[ G^\mathbb{T}(t, t^\prime) + G^>(t, t^\prime) + G^<(t, t^\prime) + G^{\tilde{\mathbb{T}}}(t, t^\prime) \right]
    \end{cases}.
\end{equation} 
Now since for $ t \neq t^\prime$, we have the relation $G^\mathbb{T}(t, t^\prime)+ G^{\tilde{\mathbb{T}}}(t, t^\prime) - G^>(t, t^\prime) - G^<(t, t^\prime) = 0 $ (which is due to the general identity of $G^\mathbb{T}(t, t^\prime) = \theta(t- t^\prime)G^{>}(t,t^\prime) +  \theta(t^\prime- t)G^{<}(t,t^\prime) $ \footnote{This identity is tautological when we consider the usual definition of time ordered Greens function, and is natural in the context of Keldysh Greens function because the $\langle \psi_+(t) \overline{\psi}_+ (t^\prime) \rangle$ component reduces to $\langle \psi_-(t) \overline{\psi}_+ (t^\prime) \rangle$  or $\langle \psi_+(t) \overline{\psi}_- (t^\prime) \rangle $ depending on the relation order between $t$ and $t^\prime$. }), we obtain
\begin{equation}
    G_{\bar{K}}(t, t^\prime) = 0  \quad \mathrm{and} \quad 
    G_{K}(t, t^\prime) =  G^{>}(t, t^\prime)   + G^{<}(t, t^\prime) \quad \mathrm{for}\quad  t \neq t^\prime. 
\end{equation}
For equal time because of the fermionic anticommutation $ G^>(t, t) -  G^<(t, t) = -i$ \footnote{This comes from $G^>(t, t) -  G^<(t, t) = -i \left[ \langle \psi_+ (t)\overline{\psi}_- (t)\rangle + \langle \overline{\psi}_+ (t)\psi_- (t)\rangle \right]  = -i \left[ \langle c c^\dagger \rangle + \langle c^\dagger c\rangle \right] = -i  $ } , and by definition, $G^\mathbb{T}(t, t) = G^{\tilde{\mathbb{T}}}(t, t) = G^>(t, t) $. Hence, we obtain for $t= t^\prime$
\begin{equation}
    \begin{cases} 
        G_{\bar{K}}(t, t^\prime) = \frac{1}{2} \left[   G^> (t, t)  - G^< (t, t) \right]  = -i/2 \\
        G_{K}(t, t^\prime)  = \frac{1}{2} \left[  3 G^>(t, t)   + G^<(t, t)   \right] = \left[ G^>(t, t)   + G^<(t, t) \right] -\frac{i}{2}. 
    \end{cases}
\end{equation} 
We thus notice that these quantities are not continuous, which can be problematic when considering local interactions in time, like in this work. \\ 

To solve this problem we aim at regularizing the Green's functions, a natural way to proceed is to set 
\begin{equation}
    G^\mathrm{reg}(t, t)= \lim_{h\rightarrow 0 }\frac{G(t, t + h)+G(t,t-h)}{2}. 
\end{equation}
Under this convention, $G_{\bar{K}}^\mathrm{reg}(t,t) = 0$ and $G_{K}^\mathrm{reg}(t,t) = G^>(t,t)  + G^<(t,t)$, thus we define  
\begin{equation}
    G^\mathrm{reg}(t,t^\prime) = G (t,t^\prime) + \delta_{t,t^\prime} \frac{i}{2} \sigma_x .
\end{equation}
 
\paragraph{The case of the interaction vertex:} We now use this observation in our case, since we are working with a Keldysh and Nambu spinor $\Phi_\mathrm{KN}$ (defined in Eq.~\eqref{eq:basis_phi_kn} ) in real space, the Green function now reads as 
\begin{equation}
    G(t,t^\prime) = \left\langle \Phi_\mathrm{KN}(t) \Phi_\mathrm{KN}^\dagger(t^\prime) \right\rangle = \left\langle\begin{pmatrix}
        \left.\begin{matrix}
            \psi_1(t) \overline{\psi}_1(t^\prime) & \psi_1(t) \psi_2 (t^\prime)\\
            \overline{\psi_2}(t) \overline{\psi}_1(t^\prime) &  \overline{\psi_2}(t) \psi_2 (t^\prime)
        \end{matrix} ~\right|
        ~
        \begin{matrix}
            \psi_1(t) \overline{\psi}_2(t^\prime) &  \psi_1(t)  \psi_1(t^\prime) \\ 
            \overline{\psi_2}(t) \overline{\psi}_2(t^\prime) &   \overline{\psi_2}(t)  \psi_1(t^\prime)
        \end{matrix} 
          \\
        \hline
        \left.
        \begin{matrix}
            \psi_2(t) \overline{\psi}_1(t^\prime)  & \psi_2(t) \psi_2 (t^\prime) \\
            \overline{\psi_1}(t) \overline{\psi}_1(t^\prime) &  \overline{\psi_1}(t) \psi_2(t^\prime)
        \end{matrix} ~\right|
        ~
        \begin{matrix}
            \psi_2(t) \overline{\psi}_2(t^\prime) &  \psi_2(t)  \psi_1(t^\prime) \\ 
            \overline{\psi_1}(t) \overline{\psi}_2(t^\prime) &   \overline{\psi_1}(t)  \psi_1(t^\prime)
        \end{matrix} 
    \end{pmatrix} \right\rangle.
\end{equation}
The regularisation found before reads as 
\begin{equation}
    G^\mathrm{reg}(t,t^\prime) = G (t,t^\prime) + \delta_{t,t^\prime} \frac{i}{2} \sigma_x \otimes \tau_z ,
\end{equation}
and to preserve the global action (with the regularised Green's function) we should satisfy
\begin{equation}
    \int \frac{\mathcal{D}\left[ \Phi_\mathrm{KN} \right]}{\mathrm{Pf}\left[G_0^{-1} \right] }~  e^{i \bar{\Phi}G_0^{-1} \Phi + i \gamma \int dt \sum_x \prod_r V_{x,r}(t) } = \int \frac{\mathcal{D}\left[ \Phi_\mathrm{KN} \right]}{\mathrm{Pf}\left[(G_0^{\mathrm{reg}})^{-1} \right] } ~ e^{i \bar{\Phi}(G_0^{\mathrm{reg}})^{-1} \Phi + i \gamma \int dt \sum_x \prod_r V^{\mathrm{reg}}_{x,r}(t) }.
\end{equation}
On simple way is to force in the limit $R \rightarrow 1$, the first order expansion of the interaction vertex to match a general form of the interaction vertex $V^\mathrm{reg}_x(t) = a + b \overline{\psi}_{1,x}(t) \psi_{2,x}(t) + b^\prime \overline{\psi}_{2,x}(t) \psi_{1,x}(t) + c \overline{\psi}_{1,x}(t) \psi_{2,x}(t) \overline{\psi}_{2,x}(t) \psi_{1,x}(t)$. By expanding $\exp(i \gamma \int dt \sum_x \prod_r V_{x,r}(t))$, and imposing an equality order by order we get for the first order in $\gamma$
\begin{equation}
    \left\langle \overline{\psi}_1 \psi_2 \overline{\psi}_2 \psi_1 \right\rangle =  \left\langle a + b \overline{\psi}_{1} \psi_{2} + b^\prime \overline{\psi}_{2} \psi_{1} + c \overline{\psi}_{1} \psi_{2} \overline{\psi}_{2} \psi_{1} \right\rangle_\mathrm{reg}
\end{equation}
where we drop $x$ and $t$ since everything is local and $\langle \circ \rangle$ denotes the Gaussian average over $G_0^{-1} $, whereas $\langle \circ \rangle_\mathrm{reg}$ denotes the Gaussian average over $\left(G_0^\mathrm{reg}\right)^{-1} $. Hence, by Wick theorem
\begin{align}
    \nonumber -G^0_{4,1}G^0_{1,4} + G^0_{4,3}G^0_{4,3} - G^0_{4,4}G^0_{3,3} &=  a -i b(G^0_{4,1} - i/2)  -i b^\prime (G^0_{1,4} - i/2)  \\
    \nonumber & - c\left[ (G^0_{4,1} - i/2) (G^0_{1,4} - i/2)   - G^0_{4,3}G^0_{4,3} + G^0_{4,4}G^0_{3,3}\right] 
\end{align}
which leads by identification to 
\begin{equation}
    \begin{cases}
        \begin{aligned}
            a - b/2 - b^\prime/2  &&=  0,  \\
              c  &&=  1,        \\
        \end{aligned}  
        \quad  
        \begin{aligned}
             c/2 - b^\prime &&=  0  \\
            c/2 - b   &&=  0      \\
        \end{aligned}    
    \end{cases}
    \quad \mathrm{resulting~in }
    \quad \begin{cases}
        \begin{aligned}
            a  &&=  \frac{1}{4},  \\
              c  &&=  1,        \\
        \end{aligned}  
        \quad  
        \begin{aligned}
             b^\prime &&=  \frac{1}{2},  \\
             b   &&=  \frac{1}{2}.     \\
        \end{aligned}    
    \end{cases}
\end{equation}
We finally obtain the interaction vertex considered in the main text 
\begin{equation}
    V^\mathrm{reg} = \frac{1}{4} + \frac{1}{2}\left[ \overline{\psi}_{1} \psi_{2} +  \overline{\psi}_{2} \psi_{1} \right] +  \overline{\psi}_{1} \psi_{2} \overline{\psi}_{2} \psi_{1} 
\end{equation}

\section{ Symplectic Structure and Pfaffian }
\label{app:symplectic}

In this Appendix, we explain why the Pfaffian emerges in the considered Gaussian integrals, and then how the symplectic structure can be used. 

\paragraph{Gaussian integrals in $\Phi_\mathrm{KN}$ basis}

Usual Gaussian Grassman integrals reads as 
\begin{equation}
    \int d\theta ~\exp\left[ - \frac{i}{2} \theta^T M \theta \right] = \mathrm{Pf} \left[ M \right]
\end{equation}
where $\theta$ is a vector of Grassman variables, and $M$ an even skewsymmetric matrix. In our case, we consider Gaussian integral of the form
\begin{equation}
    \langle Q \rangle = \int \frac{\mathcal{D}\Phi_{\mathrm{KN}} }{\mathrm{Pf}\left[-i\left(\mathbbm{1}_{R} \otimes C \right)\mathcal{G}^{-1}\right]} \exp\left[i  \frac{1}{2}\overline{\Phi}_{\mathrm{KN}}  \mathcal{G}^{-1} \Phi_{\mathrm{KN}} \right] Q.
\end{equation}
Since $\overline{\Phi}_{KN}^T = \left(\mathbbm{1}_{R} \otimes C \right)\Phi_{KN}$ (instead of simply $\overline{\Phi}_{KN}=\Phi_{KN}$), and that the field $\mathcal{G} \sim i \Phi_{KN} \overline{\Phi}_{KN}$ is respecting $\left(\mathbbm{1}_{R} \otimes C \right) \mathcal{G} \left(\mathbbm{1}_{R} \otimes C \right) = - \mathcal{G}^T $  which means that $\left(\mathbbm{1}_{R} \otimes C \right) \mathcal{G}$ is skew-symmetric, we should consider $\overline{\Phi}_{\mathrm{KN}}  \mathcal{G}^{-1} \Phi_{\mathrm{KN}}  =  \Phi_{\mathrm{KN}}^T  \left(\mathbbm{1}_{R} \otimes C \right)  \mathcal{G}^{-1} \Phi_{\mathrm{KN}} $. We thus obtain for example for $Q= V_{R}^\mathrm{reg} = \exp\left[2 \overline{\Phi}_{KN} \left( \mathbf{1}_R \otimes \sigma_x \otimes \tau_z \right) \Phi_{KN}  \right]$ the following expression 
\begin{equation}
    \langle Q \rangle =  \frac{ \mathrm{Pf}\left[-i \left(\mathbbm{1}_{R} \otimes C \right)\mathcal{G}^{-1} - 4 \left(\mathbbm{1}_{R} \otimes C \right) (\mathbbm{1}_R \otimes \sigma_x \otimes \tau_z )\right]}{\mathrm{Pf}\left[-i\left(\mathbbm{1}_{R} \otimes C \right)\mathcal{G}^{-1}\right]},
\end{equation}
and using the relation $\mathrm{det}\left[A\right] = \mathrm{Pf}\left[ A\right]^2 $, we rewrite it as 
\begin{equation}
    \langle Q \rangle  =  \sqrt{\mathrm{det}\left[ \mathbbm{1}_{4R} - i 4 \mathcal{G} (\mathbbm{1}_R \otimes \sigma_x \otimes \tau_z )\right]} ,
\end{equation}
for convenience, even if this introduce an ill defined sign that, is verify through the full computation of the Pfaffian. 

\paragraph{The Symplectic structure}
In the replica symmetric limit, or $R=1$, we are dealing with $4\times 4$ matrices. And matrices that respect symplectic relations like 
\begin{equation}
    Q = - \mathcal{C} Q^T \mathcal{C},
\end{equation}
have thus particular structure. For example, if $\mathcal{C} =\sigma_0 \otimes \tau_x $, like it is the case in basis $\mathcal{B}^\prime_{KN}$, the matrix $O$ can be parametrised as 
\begin{equation}
    Q = \begin{pmatrix}
        q_{01} & 0 & q_1 & q_2 \\
        0 & - q_{01} & q_3 & q_4 \\
        -q_4 & - q_3 & q_{02} & 0 \\
        -q_2 & -q_1 & 0 & -q_{02}
    \end{pmatrix}.
\end{equation}
By using such parametrisation, we can therefore show that 
\begin{equation}
    \frac{1}{2}\mathrm{Tr}\left[ Q \left( \sigma_z \otimes \tau_z \right) \right]^2  =   \mathrm{Tr}\left[ \left( \sigma_z \otimes \tau_0 \right) Q \left( \sigma_z \otimes \tau_0\right) \right] + 4 \sqrt{\mathrm{det}\left[Q\right]} 
\end{equation}
by computing the quantities explicitly and imposing $Q^2 = \mathbf{1}_4$.

\section{About the Causality Structure  }
\label{app:causality_struct}

In this Appendix, we argue that the shown basis $\Phi_\mathrm{KN}$ is the natural one to preserve the usual causality structure in Keldysh formalism. This was already reported in \cite{thompsonFieldTheoryManybody2023}. We reported in Appendix the relation $G^\mathbb{T}(t, t^\prime) = \theta(t- t^\prime)G^{>}(t,t^\prime) +  \theta(t^\prime- t)G^{<}(t,t^\prime)$ defined in the context of Keldysh basis only (without Nambu structure). This relation is in fact more general since for fermions, Grassman variable $\psi$ and $\overline{\psi}$ are independent, we can generally write for independent Grassman variable $\psi$ and $\phi$
\begin{equation}
\left\langle \psi_+(t) \phi_+(t^\prime) +\psi_-(t) \phi_-(t^\prime) - \psi_-(t) \phi_+(t^\prime) - \psi_+(t) \phi_-(t^\prime) \right\rangle  = 0, 
\end{equation}
hence, we notice that 
\begin{align}
    \nonumber\left\langle \vphantom{\overline{\psi}} \psi_2(t) \psi_2(t^\prime) \right\rangle & =  \frac{1}{2 } \left\langle \psi_+(t) \psi_+(t^\prime )    +    \psi_-(t) \psi_-(t^\prime )    -   \psi_+(t) \psi_-(t^\prime )  -    \psi_-(t) \psi_+(t^\prime ) \right\rangle = 0  \\
    \nonumber\left\langle  \overline{\psi}_1(t) \psi_2(t^\prime) \right\rangle & = \frac{1}{2 } \left\langle \overline{\psi}_+(t) \psi_+(t^\prime )   +    \overline{\psi}_-(t) \psi_-(t^\prime )    -   \overline{\psi}_+(t) \psi_-(t^\prime )  -    \overline{\psi}_-(t) \psi_+(t^\prime )   \right\rangle = 0 \\
    \left\langle  \psi_2(t) \overline{\psi}_1(t^\prime)  \nonumber\right\rangle & = \frac{1}{2 } \left\langle \psi_+(t)  \overline{\psi}_+(t^\prime )    +    \psi_-(t)  \overline{\psi}_-(t^\prime )    -   \psi_+(t)  \overline{\psi}_-(t^\prime )  -    \psi_-(t)  \overline{\psi}_+(t^\prime ) \right\rangle = 0 \\
    \left\langle  \overline{\psi}_1(t) \overline{\psi}_1(t^\prime)  \nonumber\right\rangle & =  \frac{1}{2 } \left\langle \overline{\psi}_+(t)  \overline{\psi}_+(t^\prime )    +    \overline{\psi}_-(t)  \overline{\psi}_-(t^\prime )    -   \overline{\psi}_+(t)  \overline{\psi}_-(t^\prime )  -    \overline{\psi}_-(t)  \overline{\psi}_+(t^\prime ) \right\rangle = 0, 
\end{align}
and only these four possibilities can arise since to get this structure, we need the minus sign in front of the minus contour. This, the considered spinor $\Phi_\mathrm{KN}$ has the right structure to preserve the usual one of the Keldysh formalism:
\begin{align}
    \nonumber \langle \Psi(t) \Psi^\dagger(t^\prime) \rangle &= i\begin{pmatrix} 
        G^A(t,t^\prime ) &  G^K(t,t^\prime ) \\
        G^{\overline{K}}(t,t^\prime ) & G^R(t,t^\prime ) 
    \end{pmatrix} \\
    &= \left\langle \begin{pmatrix}
        \left.\begin{matrix}
            \psi_1(t) \overline{\psi}_1(t^\prime) & \psi_1(t) \psi_2 (t^\prime)\\
            \overline{\psi_2}(t) \overline{\psi}_1(t^\prime) &  \overline{\psi_2}(t) \psi_2 (t^\prime)
        \end{matrix} ~\right|
        ~
        \begin{matrix}
            \psi_1(t) \overline{\psi}_2(t^\prime) &  \psi_1(t)  \psi_1(t^\prime) \\ 
            \overline{\psi_2}(t) \overline{\psi}_2(t^\prime) &   \overline{\psi_2}(t)  \psi_1(t^\prime)
        \end{matrix} 
          \\
        \hline
        \left.
        \begin{matrix}
            \hphantom{\psi_1()} 0  \hphantom{\overline{\psi}_2(t)}  &  \hphantom{\psi_1()} 0  \hphantom{\overline{\psi}_2(t^\prime)} \\
            \hphantom{\psi_1()} 0  \hphantom{\overline{\psi}_2(t)} &  \hphantom{\psi_1()} 0  \hphantom{\overline{\psi}_2(t^\prime)}
        \end{matrix} ~\right|
        ~
        \begin{matrix}
            \psi_2(t) \overline{\psi}_2(t^\prime) &  \psi_2(t)  \psi_1(t^\prime) \\ 
            \overline{\psi_1}(t) \overline{\psi}_2(t^\prime) &   \overline{\psi_1}(t)  \psi_1(t^\prime)
        \end{matrix} 
    \end{pmatrix} \right\rangle. 
\end{align}
We note that this structure no longer holds in the case of non-Hermitian evolution. This is not surprising, as the forward ($+$) and backward ($-$) evolutions are no longer identical in this context, requiring an additional degree of freedom to fully describe the dynamics.

\paragraph{Connection with usual Nambu Green's functions}

We usually define the retarded Green's function as 
\begin{equation}
    G^R (t,t^\prime) = - i \theta(t - t^\prime) \left(\begin{array}{@{}c|c@{}} \langle \{ c(t) , c^\dagger(t^\prime) \} \rangle  & \langle \{ c(t) , c(t^\prime) \} \rangle  \\ \hline
        \langle \{ c^\dagger(t) , c^\dagger(t^\prime) \} \rangle  & \langle \{ c^\dagger(t) , c(t^\prime) \} \rangle   \end{array}\right). 
\end{equation}
Lets now show that this corresponds to the retarded Green's functions we have defined previously in the Keldysh framework. We first compute the first coefficient, $-i \theta(t-t^\prime ) \langle \{ c(t) ,c^\dagger(t^\prime) \} \rangle$, which imposes $t > t^\prime$. In that context $\langle c(t) c^\dagger(t^\prime) \rangle = \langle \psi_+(t) \overline{\psi}_+ (t^\prime) + \psi_-(t) \overline{\psi}_+ (t^\prime) \rangle $ and  $\langle  c^\dagger(t^\prime) c(t)\rangle = \langle  \overline{\psi}_- (t^\prime) \psi_+(t) +  \overline{\psi}_- (t^\prime) \psi_-(t) \rangle $, thus we find that 
\begin{align}
    G^R_{11} (t,t^\prime) &=  - i \theta(t - t^\prime) \langle c(t) c^\dagger(t^\prime) \rangle \\
    &=  -i\langle \psi_+(t) \overline{\psi}_+ (t^\prime) + \psi_-(t) \overline{\psi}_+ (t^\prime) - \psi_+(t) \overline{\psi}_- (t^\prime) -   \psi_-(t) \overline{\psi}_- (t^\prime) \rangle \\
    &= -i \left\langle \psi_1(t) \overline{\psi}_1(t^\prime) \right\rangle .
\end{align}  
Similarly, we can verify the other coefficients, leading to 
\begin{equation} 
    G^R (t,t^\prime) = \begin{pmatrix}
                \psi_1(t) \overline{\psi}_1(t^\prime) & \psi_1(t) \psi_2(t^\prime) \\ 
                \overline{\psi}_2 (t) \overline{\psi}_1(t^\prime) & \overline{\psi}_2(t) \psi_2(t^\prime)
            \end{pmatrix}
\end{equation}

\section{Saddle Point Computation Details}
\label{app:saddle_point_int}

We are looking for a saddle point solution $Q^R$ \emph{local} and \emph{homogeneous} in time and space, hence the saddle point equation reads as 
\begin{equation}
    \frac{Q(t)}{2} = \left[ -i g_{0,\mathrm{KN}}^{-1} + \frac{\gamma}{2} Q \right]^{-1}_{tt,xx},
\end{equation}
by considering the Fourier transform of the right term of the equation we get
\begin{equation}
    \frac{Q}{2} =   \int_{-\infty}^\infty \frac{d\omega}{2 \pi} \int_{-\pi}^\pi \frac{dk}{2\pi}~ \left[ -i g_{0,\mathrm{KN}}^{-1} + \frac{\gamma}{2} Q \right]^{-1}_{\omega,k}
\end{equation}
because $Q$ is homogeneous we have $Q_{\omega,k} = Q V$\footnote{We have 
\begin{align}
    Q_{\omega,k} &= \int d\boldsymbol{p} d\boldsymbol{p}^\prime~ e^{ik(x-x^\prime)}e^{i\omega(t-t^\prime)} \delta(t-t^\prime)\delta(x-x^\prime) Q  =  \int_{-\infty}^\infty  \frac{d\omega}{2 \pi} \int_{-\pi}^\pi \ \frac{dk}{2 \pi} Q   = Q V
\end{align}
}, with $V$ a constant potentially diverging but that can be absorbed in $\gamma$ which is considered to be small. Since in Fourier space we have 
\begin{equation}
    g_{0,\mathrm{KN}}^{-1} = \begin{pmatrix}
       g_{R}^{-1}  & 0  \\
        0 &  g_{A}^{-1} 
    \end{pmatrix}, \quad g_{R}^{-1}  = \left(g_{A}^{-1}\right)^\dagger  =  \begin{pmatrix}
        \omega -2 J \cos(k) + h & -2i\eta \sin(k)  \\
        2 i \eta \sin(k) & \omega + 2J\cos(k) -h   
    \end{pmatrix},
\end{equation}
we can solve the retarded (or advanced) block independently. Furthermore, we note that the eigenvalues of $g_{R}^{-1}$ reads as 
\begin{equation}
   \mu_\pm = \omega \pm \sqrt{4 \eta^2 \sin(k)^2 + (h-2J \cos(k))^2 }= \omega \pm \xi_k , 
\end{equation}
and by considering that $Q^R$ and $g_{R}^{-1}$ can be diagonalised in the same basis, we obtain the equation 
\begin{equation}
    \frac{\lambda}{2} = i  \int_{-\infty}^\infty \frac{d\omega}{2 \pi} \int_{-\pi}^\pi \frac{dk}{2\pi}~ \frac{1}{\omega \pm \xi_k + i \gamma \frac{\lambda}{2} }.
\end{equation}

\paragraph{Lorentzian way:} We note that for $s>0$
\begin{equation}
   I = \int_{-\infty}^\infty  d\omega ~\frac{1}{\omega - a + is} =  \int_{-\infty}^\infty  d\omega ~ \frac{(\omega - a ) - i s }{(\omega - a)^2 + s^2 } = -i \int_{-\infty}^\infty  d\omega ~ \frac{  s }{(\omega - a)^2 + s^2 } = -i\pi, 
\end{equation}
where the real part is a boundary term that cancels, and we recognize in the imaginary part the Lorentzian which gives $\pi$. 
Hence, in our case we have
\begin{equation}
    \frac{\lambda}{2} = \int_{-\infty}^\infty \frac{d\omega}{2 \pi} \int_{-\pi}^\pi \frac{dk}{2\pi}~ \frac{\gamma \lambda/2}{(\omega \pm  \xi_k)^2 + \gamma^2 \frac{\lambda^2}{4}},
\end{equation}
obtaining the Lorentzian integral and the sign function defined as $\mathrm{sgn}(\lambda)=1$ if $\lambda > 0$ and $\mathrm{sgn}(\lambda)=-1$ otherwise
\begin{equation}
    \frac{\lambda}{2} =  \frac{ \mathrm{sgn}(\lambda)}{2 } \int_{-\pi}^\pi \frac{dk}{2\pi} = \frac{\mathrm{sgn}(\lambda)}{2},
\end{equation}
which is selfconsistent. Since the retarded block should be proportional to $\theta(t -t^\prime)$ and 
\begin{equation}
    \theta(t - t^\prime) = - \lim_{\epsilon \rightarrow 0 }\frac{1}{2\pi i} \int_{-\infty}^{\infty}d\omega~ \frac{e^{-i \omega (t-t^\prime)}}{\omega + i \epsilon }  \quad \mathrm{and} \quad 
    \theta( t^\prime - t) = \lim_{\epsilon \rightarrow 0 }\frac{1}{2\pi i} \int_{-\infty}^{\infty}d\omega~ \frac{e^{-i \omega (t-t^\prime)}}{\omega - i \epsilon } 
\end{equation}
we deduce that 
\begin{equation}
    \lambda^R =1 \quad \mathrm{and} \quad \lambda^A = - 1 
\end{equation}

\begin{figure}[h]
    \centering
    \begin{tikzpicture}
        \draw[line width=1.4pt][->] (6,-0.625) -- (2.8,-0.625) node[right] {};
        \draw[line width=1.4pt] (0,-0.625) -- (3,-0.625) node[right] {};
        

        \draw (6,-0.625)[line width=1.4pt]  arc (0:-180:3cm);
        \draw[line width=1.4pt][->] (2.9,-3.625) -- (3.2,-3.625) node[right] {};

        \draw[line width=0.6pt, dashed][->] (-1,-0.625) -- (7,-0.625) node[right] {$\mathrm{Re[\omega]}$};
        \draw[line width=0.6pt, dashed][->] (3,-4) -- (3,0.5) node[right] {$\mathrm{Im[\omega]}$}; 

        \draw[line width=1.2pt] (4.5-0.1,-2+0.1) -- (4.5+0.1,-2-0.1);
        \draw[line width=1.2pt] (4.5-0.1,-2-0.1) -- (4.5+0.1,-2+0.1);


        \node at (4.5,-1.3) [below] {$\rm pole$};
        \node at (5.5,-3.) [below] {$C_{\rm arc}$};

        \draw[line width=1pt] (0,-0.625 - 0.1) -- (0,-0.625 + 0.1) node[above left] {$-R$};

        \draw[line width=1pt] (6,-0.625 - 0.1) -- (6,-0.625 + 0.1) node[above right] {$R$};

    \end{tikzpicture}

    \caption{Scheme of the considered contour $C$, which runs as usual in a counterclockwise manner. }
    \label{fig:contour_int}
\end{figure}
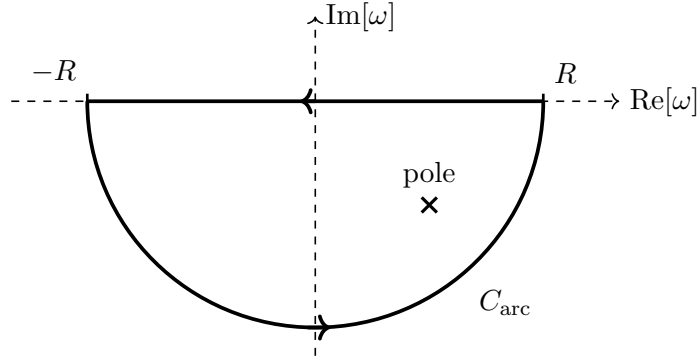

\paragraph{Residue theorem way:} This integral can also be computed through the Residue formula by considering the contour $C$ composed or the real line from $-R$ to  $R$ and an half circe in either the lower or upper half plane to close the contour (see Fig.~\ref{fig:contour_int} shown for $s>0$ in integral $I$). Hence 
\begin{equation}
   \lim_{R\rightarrow \infty }\oint_C d\omega ~ \frac{1}{\omega - a + is }  =  \lim_{R\rightarrow \infty }\int_{R}^{-R} d\omega ~\frac{1}{\omega - a + is } +  \lim_{R\rightarrow \infty } \int_{C_{\rm arc}} d\omega ~ \frac{1}{\omega - a + is } , 
\end{equation}
where by writing $\omega = R e^{i\theta}$ with $\theta \in [ -\pi,0]$, we get 
\begin{equation}
    I_\mathrm{arc} = i \int_{-\pi}^{0}  d\theta \frac{R e^{i\theta}}{R e^{i\theta} - a + is } \approx  \int_{0}^\pi  d\theta = i \pi ,
\end{equation}
where we use that $R e^{i\theta} - a + is  \approx R e^{i\theta} $ for large $R$. Moreover, since the residue of the closed contour integral is $\mathrm{Res}\left( 1/ (\omega - a + is )\right) = 1 $, we get 
\begin{equation}
    I = -2 i\pi  + i \pi =  -i\pi,
\end{equation}
we thus find the same result. We note that the computation can also be done in the upper plane where the residue is $0$. The following is similar to the previous computation. 

\section{Comparison with Lindblad in the Replica Symmetric Case} 
\label{app:lindblad}
  
In this section we perform the computation of the correlation matrix from Lindblad equation associated to our problem. The Lindblad equation describes the unconditional evolution, which correspond consider one replica, i.e. $R=1$. Hence, we should retrieve the saddle point we computed in the main text, since we are considering a replica symmetric one. We recall that the Lindbladian problem reads as 
\begin{equation}
   \frac{d \rho}{dt} = \mathcal{L}(\rho), \qquad  \mathcal{L} = -i [H, \circ] + \gamma \sum_j \left(  n_j \circ n_j - \frac{1}{2}\left\{ n_j, \circ \right\} \right) = -i\left[ H, \circ \right] -  \frac{\gamma}{2} \sum_j \left[n_j , \left[n_j, \circ \right] \right].
\end{equation}
The retarded Green's function, in the Nambu basis is defined as 
\begin{equation}
    G^R_{m,n} (t) = - i \theta(t) \left(\begin{array}{@{}c|c@{}} \langle \{ c_m(t) , c_n^\dagger \} \rangle  & \langle \{ c_m(t) , c_n \} \rangle  \\ \hline
        \langle \{ c_m^\dagger(t) , c_n^\dagger \} \rangle  & \langle \{ c_m^\dagger(t) , c_n \} \rangle   \end{array}\right) =  \left(\begin{array}{@{}c|c@{}}    G^{R,11}_{m,n}(t)    & G^{R,12}_{m,n}(t) \\ \hline
            G^{R,21}_{m,n}(t)  & G^{R,22}_{m,n}(t)   \end{array}\right),
\end{equation}
where we consider a unique time since the problem is translationnal invariant. In the Lindblad evolution, we obtain the following equation of motion for the fermionic creation and annihilation operators \cite{turkeshiDensityCurrentStatistics2024} 
\begin{equation}
    \frac{d}{dt} c_m(t) = \mathcal{L}^\dagger c_m(t) \qquad \mathrm{and}\qquad \frac{d}{dt} c_m^\dagger(t) = \mathcal{L}^\dagger c_m^\dagger(t) ,
\end{equation}
with $ \mathcal{L}^\dagger (\circ) = i[H, \circ]+ \mathcal{D}(\circ) = i[H, \circ]- (\gamma/2)\sum_j \left[  n_j \left[n_j, \circ  \right]\right] $ the adjoint Lindbladian. For a quadratic Hamiltonian $ H = \sum_{ij} h_{ij} c_i^\dagger c_j + \frac{1}{2} \sum_{ij}  \Delta_{ij} c_i^\dagger c_j^\dagger + \frac{1}{2} \sum_{ij}  \Delta_{ij}^* c_j c_i $ (with $h=h^\dagger$ and $\Delta^T = - \Delta $), we obtain  
\begin{align} 
    [H, c_m ] &= \sum_{j} -h_{mj} c_j + \frac{1}{2} \sum_{ij} \Delta_{ij} (\delta_{j,m}c_i^\dagger - \delta_{i,m} c_j^\dagger)  = - \sum_j h_{mj} c_j +\sum_i \Delta_{im} c_i^\dagger,  \\
    [H, c_m^\dagger ] &= \sum_{i} h_{im} c_i + \frac{1}{2} \sum_{ij}  \Delta_{ij}^* (\delta_{im} c_j - \delta_{jm}c_i ) = \sum_i h_{im} c_i^\dagger +\sum_j \Delta_{mj}^* c_j 
    \label{ch6:eq:hamil_commutation}
\end{align}
Then by differentiating the retarded Green's function we obtain 
\begin{align}
    \frac{d \hat{G}^R (t)}{dt} &= - i \delta(t) \begin{pmatrix}
        \langle \left\{ c_m ,c_n^\dagger \right\} \rangle &  \langle \left\{ c_m, c_n  \right\} \rangle  \\ 
        \langle \left\{ c_m^\dagger ,c_n^\dagger \right\} \rangle &  \langle \left\{ c_m^\dagger, c_n  \right\} \rangle   \\ 
    \end{pmatrix}  - i \theta(t) \begin{pmatrix}
        \langle \left\{ \mathcal{L}^\dagger [c_m] ,c_n^\dagger \right\} \rangle &  \langle \left\{ \mathcal{L}^\dagger [c_m], c_n  \right\} \rangle  \\ 
        \langle \left\{ \mathcal{L}^\dagger [c_m^\dagger] ,c_n^\dagger \right\} \rangle &  \langle \left\{  \mathcal{L}^\dagger [c_m^\dagger], c_n  \right\} \rangle   \\ 
    \end{pmatrix} 
\end{align} 
We can then use the relations of Eq.~\eqref{ch6:eq:hamil_commutation} to compute $\mathcal{L}^\dagger [c_m]$ and $\mathcal{L}^\dagger [c_m^\dagger]$, getting

\begin{align}
    & \qquad \qquad \frac{d G^R}{dt}(t) = -i \delta(t) \bold{1}_2 - i \mathbb{H}_\mathrm{eff} \hat{G}^R(t), \qquad \\
    &\mathrm{with} \qquad \mathbb{H}_\mathrm{eff} = \begin{pmatrix}
        h_\mathrm{eff} & \Delta \\
        - \Delta^* & -h_\mathrm{eff}^\dagger 
    \end{pmatrix}, \quad h_\mathrm{eff} = h - i\frac{\gamma}{2}\bold{1}_2.
\end{align} 
Hence, performing the time Fourier transform we get 
\begin{equation}
    - i \omega G^R(\omega) = -i \bold{1} -i \mathbb{H}_\mathrm{eff} G^R(\omega) \quad \mathrm{and~thus}\quad G^R(\omega) = \left(\omega - \mathbb{H}_\mathrm{eff} \right)^{-1},
\end{equation}
which in the case of the Ising chain ($h_{ij} = J\left[ \delta_{j,i+1}+\delta_{j,i-1}\right]$ and $\Delta_{ij} = \eta\left[ \delta_{j,i+1} - \delta_{i, j+1}\right]$), leads after Fourier transform  $ G^{R,ab}(q,\omega) = \sum_{n=0}^\infty e^{iqn} \left[  \hat{G}^R_{n,0}(\omega)\right]_{ab} $, with $a,b \in (1,2)$, to the result given in the main text: 
\begin{equation}
    G^R(q,\omega)^{-1} = \begin{pmatrix}
        \omega - 2J \cos(q) + i \gamma/2 &  -2 \eta i \sin(q) \\
        +2i \eta \sin(q) & \omega + 2J \cos(q) + i\gamma/2 
    \end{pmatrix}
\end{equation}

\section{NLSM Derivation}
\label{app:nlsm_deriv}

We are now in position to derive the NLSM for the monitored Ising chain.
In the action Eq.~\eqref{eq:action_full_KN_real}, the measurement term is invariant on the NLSM manifold, and only gives a constant that is irrelevant since constants should sum to zero in the limit $R \rightarrow 1$ \cite{poboikoMeasurementinducedTransitionsInteracting2025}. The term $\mathrm{Tr}\left[\Sigma \mathcal{G} \right]$ because of the non-linear constraint is also giving a constant and thus ignored. We finally focus on the fermionic determinant, and for convenience in the following we drop the $\mathrm{KN}$ subscript. Importantly, in the $\mathcal{B}^\prime_\mathrm{KN}$ (Eq.~\eqref{eq:basis_rotation}), the matrices in the action is changed (for example $\sigma_x \otimes \tau_z \rightarrow \sigma_z \otimes \tau_z $). Therefore, we have $\Sigma = \tfrac{\gamma_R}{2} \left( Q - \mathbbm{1}_R \otimes  \sigma_x \otimes \tau_z \right)$ (where here contrary to the main text we conserve the $R$ dependence in the prefactors) and $\mathcal{G} = - i Q /4 $, and we get 
\begin{equation}
    S[Q] = -\frac{i}{2}\mathrm{Tr}\left[\ln\left( -i G_{0}^{-1} + \Sigma  \right) \right] =  -\frac{i}{2}\mathrm{Tr}\left[\ln\left( -i g_{0}^{-1} +   \frac{\gamma_R  }{2} Q + \frac{\gamma - \gamma_R}{2}  (\mathbbm{1}_R \otimes \sigma_z \otimes \tau_z)   \right) \right],  
\end{equation}
where $\gamma_R = \gamma \rho_0^{R-1}$. We note that contrary to the $U(1)$ model where we can find a non trivial rotation matrix $\mathcal{R}_s$ at the replica symmetric level such that $\mathcal{R}_s \Lambda \mathcal{R}_s^{-1} = Q_0$, here we simply have $\Lambda = Q_0$. We thus define $\mathcal{R} = \mathcal{R}_R$ a rotation matrix on the replicas, and we note $G^{-1}(\gamma) = g_0^{-1} + i \tfrac{\gamma}{2} \Lambda$ with $g_0^{-1}$ the Hermitian part of the correlation matrix, and $\delta_R =  \tfrac{\gamma - \gamma_R}{2} \left(\mathbbm{1}_R \otimes \sigma_x \otimes \tau_z \right) $ which cancels for $R \rightarrow 1$, and results in 
\begin{align}
    \nonumber S[Q] &= -\frac{i}{2} \mathrm{Tr}\left[\ln\left( -i g_0^{-1} + \delta_R + \frac{\gamma_R}{2} \mathcal{R} \Lambda \mathcal{R}^{-1}   \right) \right] \\
    &= -\frac{i}{2} \mathrm{Tr}\left[\ln\left( -i   G^{-1}(\gamma_R)   +  i g^{-1}_0  -i \mathcal{R}^{-1}g^{-1}_0 \mathcal{R}   + \delta_R    \right) \right],\\
    &= -\frac{i}{2} \mathrm{Tr}\left[\ln\left( -i   G^{-1}(\gamma_R)    -i \mathcal{R}^{-1}\left[ g^{-1}_0 ,\mathcal{R} \right]  + \delta_R    \right) \right],
\end{align}
where we used that $\left[\mathcal{R}, \mathbbm{1}_R \otimes \sigma_x \otimes \tau_z \right] = 0$ and $i   G^{-1}(\gamma_R) + \tfrac{\gamma_R}{2} \Lambda = i g_0^{-1}$. Since $g_0^{-1} = i\partial_t - H_0$, we can define \footnote{We can understand this formula as follow, we write $H_0 = \sum_k \xi(k) \lvert k \rangle \langle k \rvert$, thus
\begin{equation}
    \nonumber \left[H_0, \mathcal{R} \right] =  \left(  \sum_k \xi(k) \lvert k \rangle \langle k \rvert  \right)\mathcal{R}   \left( \sum_{k^\prime}  \lvert k^\prime \rangle \langle  k^\prime \rvert \right)  -  \left(  \sum_k  \lvert k \rangle \langle k \rvert  \right)\mathcal{R}  \left( \sum_{k^\prime}  \xi(k^\prime) \lvert k^\prime \rangle \langle  k^\prime \rvert  \right)
\end{equation}
and by writing $\mathcal{R} = \sum_q e^{iqx} R_q$ we get $ \langle k \lvert \mathcal{R} \lvert k^\prime \rangle = \sum_q R_q \langle k \lvert k^\prime + q \rangle = \sum_q R_q  \delta(k,k^\prime+q)$, which gives 
\begin{align}
   \nonumber \left[H_0, \mathcal{R} \right] &=  \sum_{k,q} \left[\xi(k)- \xi(k-q) \right] \mathcal{R}_q \lvert k \rangle \langle k - q \rvert   \\
   \nonumber & \approx \sum_{k,q} \partial_k \xi(k) ~ \lvert k \rangle \langle k  \rvert \mathcal{R}_q q e^{iqx}\\
   \nonumber &= -i \hat{v} \nabla_x \mathcal{R}
\end{align} where we use $\lvert k+q\rangle = e^{iqx} \lvert k \rangle $ and where $q$ is supposed to be small to have $\xi(k+q)- \xi(k) \approx q \partial_k \xi(k)$. We obtain the relation 
\begin{equation}
    W^{(x)} = i \mathcal{R}^{-1} \left[H_0, \mathcal{R} \right] = \hat{v} \mathcal{R}^{-1} \nabla_x \mathcal{R}.
\end{equation}
} 
\begin{align}
    W &\equiv  -i\mathcal{R}^{-1} [g_0^{-1}, \mathcal{R}] =  W^{(t)}  + W^{(x)}\\
    W^{(t)} &\equiv \mathcal{R}^{-1} [\partial_t , \mathcal{R}] \approx \mathcal{R}^{-1} \partial_t \mathcal{R} \\
    W^{(x)} &\equiv i \mathcal{R}^{-1} [H_0 , \mathcal{R}]  \approx \hat{v} \mathcal{R}^{-1} \nabla_x \mathcal{R},
\end{align}
to rewrite the action as 
\begin{equation}
    S[Q] = -\frac{i}{2} \mathrm{Tr}\left[\ln\left( -i G^{-1}    \right) \right]  -\frac{i}{2} \mathrm{Tr}\left[\ln\left( 1 + i G W  + i G \delta_R   \right) \right],
\end{equation} 
with $G= G(\gamma_R) $.
 We note that both $W$ and $\delta_R$ are small and in the limit $R \rightarrow 1 $: $G(\gamma_R) \rightarrow G(\gamma)$ and $\delta_R \rightarrow 0 $. We can perform a Taylor expansion of the logarithm at second order in $W$ using the parametrisation $ G = \frac{1}{2} G^R (1+ \Lambda) + \frac{1}{2} G^A (1 - \Lambda)$ which reads as 
\begin{equation}
    S[Q] \approx \frac{1}{2} \mathrm{Tr}\left[ G W  \right]  - \frac{i}{4} \mathrm{Tr}\left[ G W GW  \right] + \Delta_R
\end{equation}
where $2 \Delta_R = \mathrm{Tr}\left[ G \delta_R  \right]  - i\mathrm{Tr}\left[ G W G \delta_R  \right]  - i\mathrm{Tr}\left[ G \delta_R   G \delta_R\right] /2 $. Since $\Delta_R $ tends to $0$ when $R \rightarrow 1$, we drop this term in the following even if the limit is only taken at the end.  


In section \ref{sec:rotation_mat_param}, we found the structure of the rotation matrix $\mathcal{R}$. For convenience we perform a change of basis, that brings this matrix in a diagonal form such that 
\begin{equation}
    \mathcal{R}^\prime = \mathcal{P}^T \mathcal{R} \mathcal{P} = \begin{pmatrix}
        \mathcal{V}_+  & 0 \\
        0 & \mathcal{V}_-
    \end{pmatrix}_K \otimes \tau_0  ,\quad  \mathcal{V}_+, \mathcal{V}_- \in SO(R)
    \label{app:eq:rotation_param}
\end{equation}
with $\mathcal{V}_+ = a+b$ and $\mathcal{V}_-=a-b$.This rotation corresponds to operate in the basis $\mathcal{B}^\prime_{KN}$, which reads as 
\begin{align}
    \Psi_{\mathrm{KN},r}^{\prime T} &= \begin{pmatrix}
        \psi_{+,x,r} & \overline{\psi}_{+,x,r} & -\psi_{-,x,r} & \overline{\psi}_{-,x,r}
    \end{pmatrix} / \sqrt{2}\\   \overline{\Psi}^\prime_{\mathrm{KN},r} &=\begin{pmatrix}
        \overline{\psi}_{+,x,r}    & \psi_{+,x,r} & \overline{\psi}_{-,x,r}& -\psi_{-,x,r}
    \end{pmatrix} / \sqrt{2}.
    \label{eq:basis_rotation}
\end{align}

In the two following paragraphs, we will evaluate the different gradient terms for the specific parametrisation $\mathcal{R}$ of Eq.~\eqref{app:eq:rotation_param}.  

\subsection{The First Order Gradients}

We start with the first order, which reads as 
\begin{equation}
    S^{(1)}\left[ Q\right] = \frac{1}{2}\left( \mathrm{Tr}\left[ G \mathcal{R}^{-1}\partial_t \mathcal{R} \right] + \mathrm{Tr}\left[ G v \mathcal{R}^{-1}\nabla_x \mathcal{R} \right] \right).
\end{equation}
\emph{Time dependent part:} We first focus on the time dependent term, and since $G$ has a trivial structure in replica space (ie. replica symmetric), we can consider $\mathrm{Tr}_R \left[ \mathcal{R}_R^{-1} \partial_t  \mathcal{R}_R \right]$ which is a matrix composed of terms $\mathrm{Tr}_R[\mathcal{V}_\pm ^{-1}\partial_t \mathcal{V}_\pm]$. Since $\mathrm{det}[\mathcal{V}_\pm] = 1 $, the derivative is null and since $ \partial_t \mathrm{det}[\mathcal{V_\pm}] = \mathrm{det}[\mathcal{V}_\pm] \mathrm{Tr}_R[\mathcal{V}_\pm ^{-1}\partial_t \mathcal{V}_\pm]$,  we find $\mathrm{Tr}_R[\mathcal{V}_\pm ^{-1}\partial_t \mathcal{V}_\pm]= 0$. Hence, we do not find first order contribution.

\emph{Space dependent term:} We have $W^{(x)}_{k,k} = \langle k \lvert W^{(x)} \rvert k \rangle = v(k) \mathcal{R}^{-1} \nabla_x \mathcal{R} $, with the velocity $v(k)$ reading as 
\begin{equation}
    v(k) = \partial_k \xi_k = \frac{8\eta^2 \sin(k) \cos(k) + 4 J \sin(k) (h - 2 J \cos(k))}{2\sqrt{4 \eta^2 \sin(k)^2 + (h - 2 J \cos(k))^2}}.
\end{equation} 
Since the first order in space can be written as 
\begin{equation}
    S[Q]^{(1,2)} =\mathrm{Tr}\left[ G v \mathcal{R}^{-1}\nabla_x \mathcal{R} \right] = \mathrm{Tr}\left[ G W^{(x)} \right] =  \int d\boldsymbol{p}~ G(\omega, k )  W^{(x)}_{k,k},
\end{equation}
where $\boldsymbol{p} = (\omega, k )$ and since $\int d\omega ~ 1/ \left[\mu^{(R/A)}_\pm (\omega,k ) \right] =  \pm i\pi$  (cf. Appendix \ref{app:saddle_point_int}) where $\mu^{(R/A)}_\pm$ are the eigenvalues of $G(\omega,k)$ the dressed Green's function. The only $k$ dependence arises in $v(k)$ which is anti-symmetric on $[-\pi, \pi]$. We obtain $\int d\boldsymbol{p} ~G(\omega,k) v(k)= 0$, thus 
\begin{equation}
    S[Q]^{(1,2)} = 0 
\end{equation}
 
\subsection{The Second Order Space Gradient }
We now consider the second order terms, and start with the spatial gradients terms, reading as 

\begin{equation}
    S[Q]^{(2,2)} = -\frac{i}{4} \mathrm{Tr}\left\{G W^{(x)} G W^{(x)}  \right\} = -\frac{i}{4} \int d \boldsymbol{p}d \boldsymbol{p}^\prime ~  \mathrm{Tr}\left\{ G(\omega,k)  W^{(x)}_{k,k^\prime} \delta_{\omega,\omega^\prime} G(\omega^\prime,k^\prime )  W^{(x)}_{k^\prime,k}  \delta_{\omega^\prime,\omega}\right\},
\end{equation}
with  $\boldsymbol{p} = (\omega, k )$ and  $\boldsymbol{p}^\prime = (\omega^\prime, k^\prime )$ and where $\delta_{\omega^\prime,\omega}$ arises because  $\langle \omega \lvert W^{(x)}_{k,k^\prime} \lvert \omega^\prime \rangle = W^{(x)}_{k,k^\prime} \delta_{\omega^\prime,\omega}$ \footnote{
This property is essentially due to 
\begin{align}
   \nonumber \mathcal{R}_t = \int_{\omega,\omega^\prime,\omega_0} ~ \lvert \omega \rangle \langle \omega \rvert e^{i \omega_0 t } R_{\omega_0} \lvert \omega^\prime \rangle \langle \omega^\prime \rvert &= \int_{\omega,\omega^\prime,\omega_0}  ~ \lvert \omega \rangle \langle \omega \rvert  \lvert \omega^\prime + \omega_0 \rangle \langle \omega^\prime + \omega_0 \rvert e^{i \omega_0 t } R_{\omega_0}   &= \int_{\omega,\omega^\prime} \delta_{\omega,\omega^\prime} \lvert \omega \rangle \langle \omega^\prime \rvert \mathcal{R},
\end{align}   
where we perform in the last equation the change of variable $\omega^\prime + \omega_0 \rightarrow \omega^\prime$.
}. We use the parametrisation $G(\omega,k ) = \left[G^R(\omega,k ) (1 + \Gamma ) + G^A(\omega,k ) (1 -  \Lambda)  \right]/2 $, therefore the term $GWGW$, contains the different product of $G^R$ and $G^A$, but only terms proportional to $G^R G^A$ are non zero due to causality. We note that $W^{(x)}$ is proportional to the identity in the Nambu sector, and assuming that $\Lambda \approx   \mathbf{1}_R \otimes \sigma_z \otimes \tau_0$, we can thus diagonalize the $G^{R/A}$ in the same basis and work with their eigenvalues that do not mix. 

Let first show that terms proportional to $G^R G^R$ or $G^A G^A$ are null, let consider 
\begin{equation}
    T^{RR} =  \int d \boldsymbol{p}d k^\prime ~  \mathrm{Tr}\left\{ G^R(\omega,k)(1+ \Lambda )  W^{(x)}_{k,k^\prime} G^R(\omega,k^\prime ) (1+ \Lambda ) W^{(x)}_{k^\prime,k}  \right\}, 
\end{equation}
since $W^{(x)}_{k,k} = \langle k \lvert W^{(x)} \rvert k^\prime \rangle = v(k) \delta_{k,k^\prime} \mathcal{R}^{-1} \nabla_x \mathcal{R} $, we get terms proportional to $ \int d\omega \int dk~ v(k)^2  \\ \mu^R_\pm(k,\omega) \mu^R_\pm(k,\omega)$ and we can show that $ \int d\omega ~  \mu^R_\pm(k,\omega) \mu^R_\pm(k,\omega) = 0 $ . 

We can now evaluate terms proportional to $G^R G^A$ which have a prefactor
\begin{equation}
   D =  \int_{-\infty}^{\infty} \frac{d \omega }{2 \pi } \int_{-\pi}^{\pi} \frac{dk}{2\pi}~ v(k)^2  \mu^R_\pm(k,\omega) \mu^A_\pm(k,\omega) =  \int_{-\pi}^{\pi} \frac{dk}{2\pi}~ \frac{v(k)^2}{\gamma} .
\end{equation}
We have in total 4 combinations proportional to $D$, thus using $\mathcal{R}^{-1} \nabla_x \mathcal{R} + \nabla_x \mathcal{R}^{-1}  \mathcal{R} = 0 $, we obtain \footnote{To go from the first equality to the second one we do the following algebra like in Ref.~\cite{kamenevFieldTheoryNonEquilibrium2011} 
\begin{align}
   \nonumber \mathrm{Tr}\left\{ (1+\Lambda) \left[ \mathcal{R}^{-1} \nabla_x \mathcal{R} \right] (1-\Lambda) \left[ \mathcal{R}^{-1} \nabla_x \mathcal{R} \right] \right\} &= \mathrm{Tr}\left\{-\nabla_x  \mathcal{R}^{-1} \nabla_x \mathcal{R}  - \Lambda  \left[ \mathcal{R}^{-1} \nabla_x \mathcal{R} \right] \Lambda \left[ \mathcal{R}^{-1} \nabla_x \mathcal{R} \right] \right\}\\
  \nonumber  &= - \frac{1}{2}  \mathrm{Tr}\left\{  \left[ \nabla_x \left(\mathcal{R}^{-1} \Lambda \mathcal{R} \right) \right]^2 \right\}
\end{align}
}
\begin{equation}
    S[Q]^{(2,2)} =-\frac{i}{4} D~ \mathrm{Tr}\left\{ (1+\Lambda) \left[ \mathcal{R}^{-1} \nabla_x \mathcal{R} \right] \right.\\ \left. (1-\Lambda) \left[ \mathcal{R}^{-1} \nabla_x \mathcal{R} \right] \right\}=  \frac{i}{8} D ~\mathrm{Tr}\left\{ \left[\nabla_x Q \right]^2 \right\} .
\end{equation}

At this stage we can repeat a similar computation for the time dependent term and thus get 
\begin{equation}
    S[Q] = \frac{i}{8} \frac{v_0^2}{\gamma} ~\mathrm{Tr}\left\{ \left[\nabla_x Q \right]^2  + \frac{1}{v_0^2}\left[\partial_t Q \right]^2 \right\} 
\end{equation}

\subsection{Non Linear Sigma Model in the Replicon Manifold}

We now implement the exact symmetry of the rotation matrix in the replica space, we thus consider $Q = \mathcal{R}_R Q_0 \mathcal{R}_R^{-1} $. As before we perform the computation for the spatial term and then deduce the temporal one. 

We note that $\nabla_x Q =   (\nabla_x \mathcal{R}_R) Q_0 \mathcal{R}_R^{-1} +   \mathcal{R}_R (\nabla_x Q_0 )\mathcal{R}_R^{-1}  +    \mathcal{R}_R Q_0 (\nabla_x \mathcal{R}_R^{-1} )$, and because of the symmetry group of $\mathcal{R}_R$, $\mathrm{Tr}_R \left[ \mathcal{R}_R^{-1} \nabla_x \mathcal{R}_R \right] = 0 $. Hence, since $Q_0$ is replica-symmetric any term where $\mathcal{R}_R^{-1} \nabla_x \mathcal{R}_R$ appears only once is null.  We get  
\begin{equation}
    \mathrm{Tr}\left\{  (\nabla_x Q)^2  \right\} =  \mathrm{Tr}\left\{ \nabla_x Q_0 \nabla_x Q_0  + V Q_0 V Q_0   + V^T Q_0 V^T Q_0 + 2 \left[\nabla_x\mathcal{R}_R^{-1} \nabla_x \mathcal{R}_R \right] Q_0^2 \right\} 
\end{equation}
with $V = \mathcal{R}_R^{-1} \nabla_x \mathcal{R}_R$. Since we are in this basis $\mathcal{B}^\prime_{KN}$, the relation previously used is modified as follow, the matrix $Q_0$ now respect the relation $ \left( \sigma_0 \otimes \tau_x\right)  Q_0^T \left( \sigma_0 \otimes \tau_x \right) = -  Q_0$, thus 
\begin{equation}
    \frac{1}{2}\mathrm{Tr}_{\mathrm{KN}}\left[ Q_0 \left( \sigma_z \otimes \tau_z \right) \right]^2  =   \mathrm{Tr}_{\mathrm{KN}}\left[ \left( \sigma_z \otimes \tau_0 \right) Q_0 \left( \sigma_z \otimes \tau_0\right) \right] + 4 \left(\sqrt{\mathrm{det}\left[Q_0\right]} \right)
\end{equation}
see Appendix \ref{app:symplectic} for the derivation. Using the standard definition in Keldysh formalism of the classical density, we have $\rho = \frac{1}{2} - \frac{1}{8} \mathrm{Tr} \left[ Q_0 (\sigma_z \otimes \tau_z) \right]$. Moreover, we can parametrize $\mathcal{R}$ such that $\mathcal{R} =  \mathcal{V}_+ T^+ + \mathcal{V}_- T^- $ with $T^{\pm} = (1\pm \sigma_z \otimes \tau_0 \otimes \mathbf{1}_R )/2$ and we obtain the useful identities :
\begin{align}
   \nonumber 2(1-\nu)  &= \frac{1}{8}   \mathrm{Tr}_\mathrm{KN}\left[Q_0 T_{zz}  \right] ^2 \\
    -2\nu & = \frac{1}{8}   \mathrm{Tr}_\mathrm{KN}\left[Q_0 T_{zz}  \right] ^2 - 2   
    \label{eq:trace_ide_q}
\end{align}
where $T_{zz} = \sigma_z \otimes \tau_z $ and $\nu = 4\rho(1-\rho)$. 
Hence, writing $V = \mathcal{V}_+^T \nabla_x \mathcal{V}_+  T^+ +\mathcal{V}_-^T \nabla_x \mathcal{V}_-  T^-$, we obtain using the previous identities Eq.~\eqref{eq:trace_ide_q}
\begin{equation}
    \nonumber\mathrm{Tr}_{\mathrm{KN}} \left\{ VQ_0 V Q_0 \right\} = \frac{8}{4} \left\{ \left[ \left(  \mathcal{V}_+ ^T\nabla_x \mathcal{V}_+\right)^2  +  \left(  \mathcal{V}_- ^T\nabla_x \mathcal{V}_-\right)^2 \right] (1-\nu) + 2\left[ \vphantom{\left(\mathcal{V}_+^T\right)^2} \mathcal{V}_+  ^T\nabla_x \mathcal{V}_+  \mathcal{V}_- ^T\nabla_x \mathcal{V}_-\right] \nu \right\} .
\end{equation} 
Since $\mathrm{Tr}_{\mathrm{KN}} \left[\left(\nabla_x\mathcal{R}_R^{-1} \nabla_x \mathcal{R}_R \right) Q_0^2 \right] = 2 \nabla_x \mathcal{V}_+^T \nabla_x \mathcal{V} + 2 \nabla_x \mathcal{V}_-^T \nabla_x \mathcal{V}_-$ and using  $\nabla_x \mathcal{V}_+ \mathcal{V}_+^T = - \mathcal{V}_+ \nabla_x \mathcal{V}_+^T $, we can write 
\begin{equation}
    \mathrm{Tr}_{\mathrm{KN}} \left[\left(\nabla_x\mathcal{R}_R^{-1} \nabla_x \mathcal{R}_R \right) Q_0^2 \right] = -2 \left(\nabla_x \mathcal{V}_+^T \nabla_x \mathcal{V}_+ + \nabla_x \mathcal{V}_-^T \nabla_r \mathcal{V}_- \right).
\end{equation}
Finally by defining $U = \mathcal{V}_+ \mathcal{V}_-^T $, we recognize the term $\nabla_x U^T \nabla_x U $, and 
\begin{equation}
   \frac{i}{8} D~\mathrm{Tr}\left\{  (\nabla_x Q)^2  \right\} =  \frac{i}{2} \nu D ~  \mathrm{Tr}_R\left\{   \nabla_x U^T \nabla_x U  \right\}. 
\end{equation}
By repeating the same computation for the temporal term, we obtain like in the main text 
\begin{equation}
    S[U]^{(2)} = \frac{2i \rho( 1- \rho)}{\gamma}  \mathrm{Tr}_R\left\{   v_0^2 \nabla_x U^T \nabla_x U  + \partial_t U^T \partial_t U  \right\}.
\end{equation}

\bibliography{mabiblio}
\end{appendix}

\end{document}